\documentstyle[aps,pre]{revtex}
\begin{document}
\draft
\title{A domain wall between single-mode and bimodal states and its
transition to dynamical behavior in inhomogeneous systems}
\author{M. van Hecke $^1$ and B. A. Malomed $^2$}
\address{$~^1$ Instituut-Lorentz, University of Leiden,\\
P.O. Box 9506, 2300 RA Leiden, the Netherlands,\\
e-mail martin@lorentz.leidenuniv.nl 
\footnote{Address from 1 Oct. 1996: The Niels Bohr Institute,
Blegdamsvej 17, 
2100 Copenhagen \O, Denmark}
\\
$~^{2}$ Department of Applied Mathematics,
School of Mathematical Sciences,\\ Raymond and Beverly Sackler Faculty of
Exact Sciences, Tel Aviv University, Ramat Aviv, Tel Aviv
69978, Israel\\ e-mail malomed@leo.math.tau.ac.il}

\date{\today}
\maketitle
\begin{abstract}
We consider domain walls (DW's) between single-mode and bimodal states
that occur in coupled nonlinear diffusion (NLD), real Ginzburg-Landau
(RGL), and complex Ginzburg-Landau (CGL) equations with a spatially
dependent coupling coefficient.
Group-velocity terms are added to the NLD and RGL equations, which breaks
the variational structure of these models.
In the simplest case of two coupled NLD equations, we
reduce the description of stationary configurations to a single second-order
ordinary differential equation. We demonstrate analytically that a 
necessary condition for existence of a 
stationary DW is that the group-velocity must be
below a certain threshold value. Above this threshold, dynamical behavior
sets in, which we consider in detail. In the
CGL equations, the DW may generate spatio-temporal chaos,
depending on the nonlinear dispersion. A spatially dependent coupling 
coefficient as considered in this paper
can be realized at least in two different convection systems: a
rotating narrow annulus supporting two traveling-wave wall modes, and a
large-aspect-ratio system with poor heat conductivity at the lateral 
boundaries, where the two phases separated by the DW
are rolls and square cells.
\end{abstract}

\pacs{47.20.Ky, 03.40.Kf, 75.60.Ch}
\vspace{1cm}

\section{Introduction}

Close to a pattern-forming transition, many hydrodynamic systems can be 
described by coupled one-dimensional (1D) Ginzburg-Landau equations for 
some envelope functions (order parameters) $A$ and
$B$. These equations were first introduced in an explicit form (with 
purely real coefficients, but also with the
group-velocity terms) by Cross \cite{Cross}. In the presence of a small
over-criticality parameter, and when rewritten in terms 
of two real amplitudes and two real phases, these equations for two complex
order parameters are completely tantamount to a single fourth-order complex 
equation (the ``complex Swift-Hohenberg'' equation) introduced still earlier
in Ref. \cite{Zphys}. As is well known, the coefficient in front of the 
cross-coupling (CC) term is an essential parameter of these equations.
For a small CC coefficient, homogenous bimodal states of the form $|A|=|B|$ 
occur, while for sufficiently large CC, this  bimodal state is unstable,
and the system evolves into a single-mode
state in which either $A$ or $B$ is zero \cite{Coullet}. 
In this work, we
consider the case that the CC coefficient is a function of
the spatial coordinate $x$. In particular, we focus on the
case that the CC coefficient 
is slightly above $1$ for $x\rightarrow -\infty$
and slightly below $1$ for  $x\rightarrow \infty$.
The simplest state that can occur in this case is a spatial
juxtaposition of a single and a bimodal state, separated by a so-called
domain-wall.
We study this domain-wall in the NLD, RGL, and CGL equations.
We show that non-variational effects can destroy
the stationary domain-wall and study the dynamical states that occur in 
this
case. In particular we focus on the effect of the group-velocity terms
in the coupled equations.
 
This paper is organized as follows. First we introduce a simplified
model for the occurrence of domain-walls. For stationary solutions of
this model we derive a perturbation equation that allows us to
predict the vanishing of stationary domain-walls when the
group-velocity is sufficiently large. This is confirmed by
numerical simulations of this model, and we describe the ensuing
dynamical states.
We study the relevance of this model for coupled RGL and CGL equations
by performing numerical simulations. In particular for the CGL equations,
a broad spectrum of dynamical behavior is observed, including
spatio-temporal chaos. It is very difficult to characterize these states,
and we restrict ourself to a limited exploration of the possible
dynamical states.

Some of
the essential features of the CGL equations 
that are important in the study
of the domain-wall, i.e., the
stability of the single and bimodal states as a function of the CC 
can also be studied in simpler models.
We will first consider two coupled nonlinear diffusion (NLD) equations,
to which we add group-velocity terms borrowed from the CGL equation 
that break the variational representation of the NLD equations (however, 
the group-velocity terms should not be added in the above-mentioned
case of the Rayleigh-B\'enard convection in the
inhomogeneous large-aspect-ratio system
\cite{MNT}). 
In the simplest approach, we take the amplitudes real-valued:
\begin{mathletters}
\begin{eqnarray}
\partial_t A  + c \partial_x A &=& A + \partial_x^2 A -
(A^2 + g(x) B^2) A ~,\\
\partial_t B  - c \partial_x B &=& B + \partial_x^2 B -
(B^2 + g(x) A^2) B ~,
\end{eqnarray}
\end{mathletters}
These NLD equations are the simplest set of equations that admit
the DW's that we are interested in.
These equations have no direct physical interpretation in terms of
amplitude equations, but they have the advantage over more realistic models
that the effect of the group-velocity terms on the domain-walls can be 
studied analytically.
Since the group-velocity terms have opposite sign for the $A$ and
the $B$ equation of the coupled CGL equations (\ref{cgl}),
we retain this property in the NLD equations.
Without the cross-coupling, the group-velocity terms
generate a counter propagation of the patterns in $A$ and $B$.
When the equations for $A$ and $B$ are coupled, the effect of
the group-velocity terms can not be given in terms of a simple propagation
 rule.

We will derive, in a certain limit, a single ordinary differential
equation for the order parameter $\chi := \tan^{-1}(A/B)$ 
that describes
stationary configurations. Using this 
equation, we will be able to investigate the existence of a stationary DW
analytically. In section \ref{varisec}
we consider the case that the group-velocity terms are
absent. The equations are then variational and for the particular case
that the CC coefficient, as a function of the
spatial coordinate $x$, is proportional to ${\rm tanh}(\kappa x)$,
an exact
analytical solution for the DW can be obtained. Next, we
extend in section \ref{nldevsec}
the equation for $\chi$ to include the non-variational
group-velocity terms, and when we take $g(x)$ to be a
step function, 
an analytical solution can be constructed. Analyzing the latter solution,
we find a transition from a regime where the DW
is stationary to a regime where no stationary DW exists.
This transition, which we show to have a nice geometrical interpretation,
occurs when the group-velocity $c$ is equal to the
critical value $c_{crit}:= \pm 2\sqrt{|g-1|}$.
We will show
that this result remains valid for a broad class of inhomogeneous
CC coefficients.
 
Direct simulations of the coupled NLD
equations, presented in section \ref{secFKnum}, confirm the analytical 
results and
show that there is a value of the group-velocity, $c_i$, where
the stationary domain-wall looses its stability.
This instability
occurs because the position of the domain-wall diverges when 
$c$ approaches $c_{crit}$. 
The difference between $c_{crit}$ and $c_i$ is of the order of
a few percent.
For a group-velocity just beyond $c_i$, 
the DW is seen to perform small-amplitude chaotic oscillations around a
mean position, without essentially disturbing the DW's shape.

The study of the coupled CGL equations is of a more explorative nature.
As a first step towards the study of the full coupled CGL equations,
we allow the amplitudes $A$ and
$B$ to be complex valued. These equations are reminiscent of the RGL
 equations,
but it should be noted that in physical applications, the RGL equations 
have no
group-velocity terms;  nevertheless
we refer to this model as the coupled RGL equations.
We explain that the main properties of
the stationary domain-walls in this model can be described by
the NLD equations.
Simulations of coupled RGL equations, presented in
section \ref{rglsec}, show a transition between the stationary
domain-walls and a dynamical state with the increase of the group-velocity
that is similar to the NLD-case.
The role of wavenumbers of the initial conditions,
which constitutes the main feature that is not present
in the NLD equations, is discussed briefly.
 
For the coupled CGL equations, the numerical results presented
in section \ref{cglsec} show that the nonlinear dispersion terms 
may render
the domain-walls unstable, even
when the group-velocity terms are absent. Spatio-temporal
disordered states often occur.
The oscillations of the DW can be understood qualitatively
to arise from the large gradients of $|A|$ and $|B|$ around the 
domain-wall.
                                          
It may be pertinent to note that both complex and real Ginzburg-Landau
equations with the coefficients in front of the {\it linear} terms smoothly
depending upon the spatial coordinate have been the subject of many
studies \cite{Malomed1,Malomed2,inhomo}.
It was found that a parameter ramp in the real or complex GL equation
can perform wavenumber selection \cite{Malomed1,inhomo}. In the case of
the full CGL equation, it can also render the single-mode traveling waves 
unstable, and can trap subcritical solitary pulses \cite{Malomed2}.
However, as far as we
know, there have been no studies aimed to consider effects of a spatial
dependence in the coefficient in front of the nonlinear CC term.

A spatially dependent CC can be realized in Rayleigh-B\'enard 
convection in a rotating annulus of non-constant width \cite{Hecke}.
Convection in rotating systems has recently
been the focus of several studies \cite{Hecke,Zhong,Ning,Goldstein,Kuo},
and quasi-1D traveling waves were shown to occur in these
systems near the vertical side-walls of the annulus.
In a rotating annulus, there are two so-called wall modes,
localized, respectively,
near its inner and outer side walls. Note that there is only one wall-mode 
per side-wall (instead of two); this is due to a symmetry breaking that
is induced by the rotation \cite{Hecke,Zhong,Ning,Goldstein,Kuo}. 
The amplitude
equations describing slow modulations of these modes in the
co-rotating reference frame are two coupled
cubic complex Ginzburg-Landau (CGL) equations \cite{Hecke}. The
strength of the CC between the two wall modes sensitively depends
on the width of the annulus.
When the annulus is not uniform but has a varying width, the corresponding
CC can be made to vary across its
critical value as a function of the longitudinal spatial coordinate (which is
going along the circumference of the rotating annulus). Other coefficients of
the amplitude equations describing such a system will also
depend on this coordinate, but if we assume that these coefficients,
in contrast to the CC coefficient, are not close to a critical value,
their spatial dependence may be ignored, provided that it is smooth
enough.
 
A similar problem may be implemented in a related but
different physical system, viz., the Rayleigh-B\'enard 
convection in a (non-rotating) large-aspect-ratio
cell. In this case, the two modes are two orthogonal sets of parallel
rolls. A single-mode state is stable provided that an effective CC coefficient
between the orthogonal rolls is larger than a certain
minimum value, while in the opposite case a square-lattice
pattern (the bimodal
state obtained as a superposition of two orthogonal sets
of rolls) is stable \cite{Busse}. 
Usually, the actual value of the CC coefficient is well above the 
above-mentioned minimum,
so that the square lattice is unstable. However, in special
cases, e.g., for the convection between horizontal surfaces with poor heat
conductivity, the CC coefficient may fall below the minimum \cite{Siv}. 
One may construct a suitable inhomogeneous system, for example, by
means of a variable-thickness lid put on top of the convection layer,
such that the local CC coefficient, being a function of the spatial
coordinates, is passing through the minimum value. Then one may expect a 
stationary DW separating the rolls and square lattice, which is 
impossible in the homogeneous system \cite{MNT,hoyle}. 

\section{The variational case}              \label{varisec}

In this section we will focus on the coupled NLD
and RGL equations, which are
the simplest sets of equations where a DW
between a single-mode and a bimodal state can occur.
The dynamics of these systems without the group-velocity terms are
relaxational in the sense that a
Lyapunov functional 
${\cal L}$ 
exists for each of them, allowing to predict
final states by minimizing it.
We derive a perturbation theory for a weak inhomogeneity in 
the cross-coupling coefficients 
of the coupled
NLD equations, that allows us to find a closed-form expression
for the DW for a special choice of the inhomogeneity.
This perturbation theory will be a starting point
in the next subsection, where we will focus on the group-velocity terms that 
make the equations non-variational.

The equations that we consider in this section are
\begin{mathletters}\label{RGLstat}
\begin{eqnarray}
\partial_t A  &=& A + \partial_x^2 A -(|A|^2 + g(x) |B|^2) A ~,\\
\partial_t B  &=& B + \partial_x^2 B -(|B|^2 + g(x) |A|^2) B ~,
\end{eqnarray}
\end{mathletters} 
where $A$ and $B$ are real-valued in the NLD case, and complex in the RGL
case. 
The critical value of $g$ is $1$: for
$g<1$ the bimodal state $|A|=|B|=\sqrt{1/(1+g)}$ is stable,
whereas for $g>1$ this bimodal state loses its stability and
the single-mode state with $|A|=1$ and $B=0$ (or vice
versa) becomes stable \cite{Coullet}.
The CC coefficient $g(x)$ is assumed to decrease monotonically from 
slightly above to slightly below its critical value as a 
function of the spatial coordinate $x$. 
In what follows below, we will set $g(x):= 1 + \gamma(x)$, where
$\gamma(x)$ is a small monotonically decreasing function
of the spatial coordinate $x$, such that $\gamma(x)$ is positive at
$x<0$ and negative at $x>0$. We assume that $\gamma (x)$ saturates at
$x\rightarrow \pm \infty$, i.e., it assumes certain asymptotic values
$\gamma(-\infty)=\gamma _{{max}}>0$, and
$\gamma (+\infty)=\gamma _{{min}}<0$. 
The solution that one expects for this choice of $g(x)$
is a stationary DW located  around $x=0$ which matches the single-mode
and bimodal states existing, respectively, for negative and positive $x$.

The coupled RGL equations can be derived from the Lyapunov functional 
\begin{equation}
{\cal L} = \int dx \left\{ |\partial_x A|^2 +|\partial_x B|^2 - 
(|A|^2 + |B|^2) +\frac{1}{2}(|A|^4 + |B|^4) +
 (1+ \gamma(x)) |A|^2 |B|^2 \right\} 
\end{equation}
by setting $\partial_t A = - {\cal \delta L}/{\cal \delta} A^*$ and 
$\partial_t B = - {\cal \delta L}/{\cal \delta} B^*$.
As is commonly known, the Lyapunov functional may only decrease in 
time, and, as it is bounded from below, a final state corresponds to a 
minimum of ${\cal L}$. The final
state that corresponds to a global minimum of ${\cal L}$ has a zero
wave number, which suggests to consider also the particular case of real
$A$ and $B$. Thus one obtains the coupled NLD equations.
For finite systems with periodic boundary conditions, the RGL equations
may evolve to stationary states with nonzero wavenumber. This
is discussed in section \ref{rglstatsec}.

We will now focus on the stationary solutions of the NLD
equations, and
set the derivatives with respect to time  equal to zero.
The ensuing equations can be written as four coupled ordinary 
differential
equations for the real-valued 
$A$, $\partial_x A$, $B$ and $\partial_x B$.
When we identify
$A$ and $B$ with position coordinates and their derivatives 
with generalized momentum coordinates, the equations for a 
stationary domain-wall can
be written as the {\em Hamilton equations} corresponding to the
Hamiltonian
\begin{equation}
H= \frac{1}{2} \left\{ (\partial_x A)^2 +(\partial_x B^2)+ (A^2 + B^2)
 -\frac{1}{2} (A^4 + B^4)- (1+\gamma(x)) A^2 B^2 \right\} ~, \label{H}
\end{equation}
where $x$ is the ``time'' coordinate. 
Although the Hamiltonian is similar to the Lyapunov functional,
the difference in the signs that occur between these expressions should be
noted.
%

In general we can not solve the Hamilton equations for arbitrary
$\gamma(x)$. To proceed we will develop a perturbation expansion by
taking $\gamma_{{ min}}$ and $\gamma _{{ max}}$ small. 
We will complement this condition by the assumption that
$A(x)$ and $B(x)$ are slowly varying functions of $x$, so that the diffusive
terms in Eqs. (\ref{RGLstat}) are small in comparison with the other terms 
Below we will determine, in a self-consistent way,
that the spatial scale $L$ over which $A$ and $B$ vary is
of order $1/\sqrt{|\gamma|}$.
When the diffusive terms and the coupling term $\gamma(x) A^2 B^2$
of the Hamiltonian (\ref{H}) are small, it follows that $(A^2 + B^2)$
is almost constant \cite{MNT}.
This suggests the following representation for
$A(x)$ and $B(x)$ \cite{MNT}:
\begin{equation}
A(x)=R(x)\, \cos\chi(x);\; B(x)=R(x)\, \sin\chi(x).\label{chi}
\end{equation}
The
single-mode states are those with $\chi$ being an 
integer multiple of $\pi/2$, and bimodal states have
$\chi =  \pi/4 + n\pi/2$ (throughout this paper, $n$ will represent
an integer).
%
%
When we substitute this representation into the Hamiltonian (\ref{H})
we obtain
\begin{eqnarray}
H= R^2-\frac{1}{2} R^4 + R'^2+ R^2 \chi'^2- 4 R R' \chi' \cos(\chi) 
\sin(\chi)
\nonumber \\-\gamma(x) R^4 \cos(\chi)^2  \sin(\chi)^2~, \label{hchi}
\end{eqnarray}
where a prime denotes differentiation with respect to $x$.
Stationary solutions of the coupled NLD equations
are found by determining the minima of the Hamiltonian (\ref{H}).
Since we assumed that $A$ and $B$ are slowly varying, and that $R$ is
almost constant, $R' \ll \chi'$
and in the simplest non-trivial approximation we may set $R$ equal to one \cite{MNT}.

%
%
Differentiating the Hamiltonian (\ref{hchi}) with respect to $x$
and neglecting higher order terms,
we obtain the following perturbation equation:
\begin{equation}
\chi''(x) = \frac{1}{4} \gamma(x) \sin(4\chi),\label{chi''}
\end{equation}
where the prime stand for $d/dx$. It follows from here that the
relation between the small quantity $\gamma$ and
a large scale $L$ of variation of the function $\chi(x)$, which was implicitly
assumed above, is $L^{-2}\sim \gamma$ \cite{MNT}.

As an example, we may solve  
Eq. (\ref{chi''}) exactly for the special choice 
$\gamma(x)=  -\kappa^2 \tanh(\kappa x)$. The solution is then
\begin{equation}
\chi_0(x) = \frac{1}{2} \tan^{-1} \left(e^{\kappa x}\right),\label{exact}
\end{equation}
which can be verified by substitution;
note the scaling of $x$ with $\sqrt{|\gamma|}$.
This choice for $\gamma(x)$ and
the corresponding solution are shown 
in Fig. 1 for $|\gamma| =0.2$.
We will proceed by investigating whether this type of the DW
solutions exist for more general $\gamma(x)$ and when the group-velocity
terms are included.

\section{The NLD equations with the group-velocity}\label{nldevsec}

In this section we will 
investigate  what happens to the domain-walls when 
the group-velocity terms that are similar to those in the CGL equation
are included in the NLD equations. 
These terms destroy
the variational structure of the NLD equations, hence their 
final states need no longer be stationary. 
We will show that stationary DW's cannot exist when 
the group-velocity is above a certain critical value.
Numerical simulations of the NLD
equations corroborate this prediction, and show 
that, beyond the threshold, time-periodic or disordered states occur.

The NLD equations with the group-velocity terms are
\begin{mathletters}\label{FK}
\begin{eqnarray}
\partial_t A  + c \partial_x A &=& A + \partial_x^2 A -
(A^2 + g(x) B^2) A ~,\\
\partial_t B  - c \partial_x B &=& B + \partial_x^2 B -
(B^2 + g(x) A^2) B ~,
\end{eqnarray}
\end{mathletters}
As was demonstrated in Ref. \cite{Malomed3}, one can 
use a {\em balance equation} for the
Hamiltonian (\ref{H}) to treat effects of the
small group-velocity terms on the stationary
solutions. Differentiating (\ref{H}) and making use of the stationary 
version of Eqs. (\ref{FK}), one finds
\begin{equation}
\frac{dH}{dx}= c\left[ (\partial_x A)^2 - (\partial_x  B)^2\right] ~.
\label{eqwithc}
\end{equation}
Using the ``polar'' representation defined by Eqs. (\ref{chi})
yields an effective equation for $\chi$, which is a generalization of 
Eq. (\ref{chi''}) 
(the same equation, albeit for constant $\gamma$, was obtained in Ref. [13]):
\begin{equation}
\chi''(x) - \frac{1}{4}\gamma(x)\, \sin(4 \chi) + c \chi' \cos(2
\chi)=0.\label{balance}
\end{equation}
This equation is the 
basis for the perturbative analysis presented below.

It should be noted that this equation is invariant under a scale 
transformation
$\gamma\rightarrow \delta \gamma$, $x\rightarrow x/\sqrt{\delta}$ and
$c \rightarrow \sqrt{\delta} c $. This freedom can in principle
be used to scale out $c$, but we will not do this; this scale-invariance
is reflected, however,
in the formula for $c_{crit}$ that is obtained below.

\subsection{Phase-space analysis}\label{secFKana}

It will be convenient to rewrite Eq. (\ref{balance}) in the
form of a two-dimensional {\it non-autonomous} dynamical system:
\begin{mathletters}\label{dynamical}
\begin{eqnarray}
\frac{d\chi}{dx} &=& \psi ~, \\
\frac{d\psi}{dx} &=& - c~ \psi \cos (2 \chi) + \frac{1}{4} \gamma(x)
\sin(4 \chi)~.
\end{eqnarray}\end{mathletters}
Fixed points of Eqs. (\ref{dynamical}) are
$(\chi=n \pi/4, \psi=0)$, which correspond to single-mode stationary
solutions of equations of Eqs. (\ref{FK}) at even $n$, and to
bimodal solutions at odd $n$.
Hetero-clinic orbits going from one fixed point
at pseudo-time $x=-\infty$ to another fixed point at
$x=\infty$ correspond to the DW's that we are interested in.
Since we have chosen $\gamma(x)$ to be a decreasing function of $x$,
these orbits go from a fixed point with even $n$
to one with odd $n$. 

The system (\ref{dynamical}) is invariant with respect to the following
symmetry transformations:
\begin{itemize}
\item[(i)] $\chi \rightarrow \chi + n \pi$,
\item[(ii)] $\chi \rightarrow \chi + \pi/2$, $c \rightarrow -c$.
\item[(iii)] $x\rightarrow-x$, $\psi\rightarrow-\psi$, 
$c \rightarrow -c$, $\gamma(x)\rightarrow\gamma(-x)$.
\end{itemize}
With regard to these symmetries, it is sufficient to consider only the
hetero-clinic orbit that goes from the fixed point $(\chi,\psi)=(0,0)$, to
be referred to as FP0, to the bimodal fixed point
$(\chi,\psi)=(\pi/4,0)$, which will be called
FP1; the other DW's can be obtained by applying a combination of the
transformations (i), (ii) and  (iii) to this hetero-clinic orbit. 


An essential feature of the dynamical system (\ref{dynamical})
is that the direction of the phase-flow in the $(\chi,\psi)$ 
plane is not
fixed, but depends on the current value of $\gamma(x)$.
Therefore, even when the functional form of
$\gamma$ is fixed, the possible orbits
of equation (\ref{dynamical}) through a certain point 
form, in general, a one-parameter family.

However, when we take $\gamma(x)$ to be a step function, i.e.,
\begin{mathletters}\label{step}
\begin{eqnarray}
\gamma(x)&=&\gamma_{{max}} >0,\; x<0~,\\
\gamma(x)&=&\gamma_{{min}} <0,\; x> 0~,
\end{eqnarray}\end{mathletters}
there are only 2 orbits through a certain point; 
one for $\gamma=\gamma_{max} $
($x<0$) and one for $\gamma=\gamma_{min}$ ($x>0$).
Notice that the lack of continuity of $\gamma(x)$
does not contradict the assumption that the stationary solution 
for $\chi$ is a
smooth function of $x$; we will see below that solutions corresponding
to this discontinuous $\gamma (x)$ are smooth indeed.
In fact, the scaling properties of equation (\ref{balance}), that also
hold for the dynamical system (\ref{dynamical})
yield that when we rescale $\gamma\rightarrow \delta \gamma$, with
$\delta<1$, the function $\gamma(x)$ becomes effectively steeper due
to the rescaling of the spatial-coordinate; when
$\delta \downarrow 0$, $\gamma(x)$ becomes, in a sense, infinitely 
small and infinitely steep.

We are now interested in the behavior of the hetero-clinic orbits as a 
function of $c$.
A hetero-clinic trajectory corresponding to the DW exists provided that the
the outgoing (unstable) manifold of FP0, which
we will refer to as $W_0^{{out}}$, intersects the ingoing (stable) manifold 
of FP1, to be referred to as $W_1^{{in}}$
(see Figs. \ref{fig2} and \ref{fig2sub}). 
Obviously, $W_0^{{(out)}}$ and $W_1^{{(in)}}$
have to be calculated, respectively, for $\gamma =\gamma_{{max}}$ and
$\gamma= \gamma_{{min}}$. The question of existence
of a domain-wall has thus been reduced to checking whether 
$W_0^{{(out)}}$ and $W_1^{{(in)}}$ intersect.
The point of intersection then gives the values
of $\chi$ and $\psi$ at $x=0$.

This is illustrated in the Figs. \ref{fig2} and \ref{fig2sub}, where
we have plotted, for $\gamma_{min}=-0.1$ and $\gamma_{max}=0.1$,
these two manifolds for a range of values of the group-velocity.
The point of intersection has by definition $x=0$, and so from
these Figs. we can read off the values of $\chi$ and $\chi'$ at 
$x=0$ as a function of $c$.
The point of intersection of $W_0^{{(out)}}$ and $W_1^{{(in)}}$
shifts towards FP0 when $c$ is decreased.
As a consequence, the location of the domain-wall, i.e, 
the value of $x$
where $\chi =\pi/8$, shifts to larger and larger values.
In fact, when $c$ approaches a  certain value which we define as 
$c_{crit}$, the intersection of
 $W_0^{{(out)}}$ and $W_1^{{(in)}}$ approaches FP0 and
the location of the domain-wall, that corresponds to
the hetero-clinic orbit, diverges to infinity 
(see Fig. \ref{statfig}).
 
Below we will find that $c_{crit}=-2\sqrt{|\gamma_{min}|}$.
Note that the square-root dependence follows from the scaling 
properties;
only the numerical factor $-2$ needs to be determined.
There are two complementary methods for the
determination of the critical velocity. The first method involves
analytically solving the flow-line
equations of the dynamical system (\ref{dynamical})
and the second method involves an inspection
of the geometry of the flow-lines.
After a description of these two methods we discuss the validity of
$c_{crit}$ for inhomogeneities that are not of the form (\ref{step}).

\subsubsection{Analytic expression for the flow-lines.}
The manifolds $W_0^{{(out)}}$ and  $W_1^{{(in)}}$
 can be found exactly when $\gamma$ is piecewise
constant, as in Eqs. (\ref{step}), 
because we can solve the equation for the 
trajectories of the dynamical system (\ref{dynamical}) explicitly 
when $\gamma$ is a constant. This equation
is obtained by dividing Eq. (\ref{dynamical}b) by
Eq. (\ref{dynamical}a) 
which yields for constant $\gamma$
:
\begin{equation}
\frac{d\psi}{d\chi} = -c \cos (2\chi) +\frac{\gamma}{4\psi} \sin (4 \chi).
\label{floweq}
\end{equation}
To solve equation (\ref{floweq}), we first perform a coordinate
transformation 
by introducing $\zeta:=
\frac{1}{2} \sin(2 \chi)$, which yields
\begin{equation}
\frac{d\psi}{d\zeta} = -c + \gamma \frac{\zeta}{\psi}~.\label{homo}
\end{equation}
Eq. (\ref{homo}) is homogeneous, and by
defining $\lambda = \psi/\zeta$ and some rewriting,
we find
\begin{equation}
\frac{1}{\zeta} d\zeta = \frac{1}{-c + \gamma/\lambda - \lambda} 
d \lambda~.
\label{sepeq}
\end{equation}
The manifold $W_0^{{(out)}}$ corresponds to the case that $\lambda$
is a constant, which occurs for $-c + \gamma/\lambda - \lambda=0$.
By transforming this solution back to the $\chi,\psi$ coordinates
we obtain for $W_0^{{(out)}}$:
\begin{equation}
\psi = \lambda_0/2 \sin(2 \chi)~, \label{sol0}
\end{equation}
where $\lambda_0$ satisfies  $-c + \gamma/\lambda_0 - \lambda_0=0$,
and $\gamma=\gamma_{max}$.

The manifold $W_1^{{in}}$ is found by a straightforward
integration of equation (\ref{sepeq}),
which yields the unpleasant equation
\begin{equation}
\ln(\zeta) = \frac{c~ \mbox{atanh}\frac{c + 2\lambda}
{\sqrt{4 \gamma+ c^2}}}{\sqrt{4 \gamma + c^2}}
- \frac{\ln(-\lambda+ c \lambda + \lambda^2)}{2} + K~,\label{sol1}
\end{equation}
where $K$ is an arbitrary constant of integration,
and $\gamma=\gamma_{min}$. $K$ has to be 
chosen so that the trajectory is passing through the point FP1, where 
$(\zeta,\lambda)=(\frac{1}{2},0)$. Although we do not have explicit expressions 
for the manifolds in terms of the original variable $\chi$,
we can find from Eqs. (\ref{sol0}) and (\ref{sol1}) under
which conditions the manifolds
$W_0^{{out}}$ and $W_1^{{in}}$ intersect. The intersection point 
can be found by
substituting $\lambda=\lambda_0$ into Eq. (\ref{sol1}).

Simply solving the ensuing equation numerically yields that
$W_1^{{(in)}}$ and $W_0^{{(out)}}$ intersect only (apart from
the intersection at FP0 itself)  when
\begin{equation}
c>c_{{crit}}:= -2\sqrt{-\gamma_{{min}}},\label{condition}
\end{equation}
where it should be noted that $c$ is assumed to be negative.

This means that if the group-velocity $c$, which
is amenable for the non-variational effects in this model, is below
the threshold $-2\sqrt{-\gamma_{min}}$, there can not 
be a stationary DW. 
It should be noted that we have focused here on the domain-wall
that has a single-mode state for negative $x$ and a bimodal state
for positive $x$.
When we reflect $x$, and study a domain-wall of opposite chirality,
i.e., going from a bimodal state at $x\rightarrow
-\infty$ to a single-mode state at $x\rightarrow +\infty$,
according to symmetry (iii) the critical velocity changes sign.
When $\gamma(x)$ goes from negative values for $x<0$ to positive values
for $x>0$, the ensuing
stationary domain-walls can not exist for sufficiently large 
{\em positive} $c$.
We will  encounter domain-walls of both chiralities in the numerical
simulations presented below.


\subsubsection{Geometrical interpretation.}
A simple geometric interpretation of the threshold is illustrated in the
Figs. \ref{fig2} and  \ref{fig2sub}.
As is illustrated in the phase-portraits
(see Figures \ref{fig2}a and \ref{fig2}b),
$W_1^{{(in)}}$ intersects $W_0^{{(out)}}$ for all positive $c$,
and therefore we will
concentrate now on the case of negative $c$.
The central point is that
for negative $c$ and negative $\gamma$, FP0 is a spiral when
$c>c_{crit}$, and a
saddle when $c<c_{crit}$. Therefore, the motion of $W_1^{{(in)}}$,
for which $\gamma=\gamma_{min}<0$, in the neighborhood of FP0
is spiral-like when $c>c_{crit}$;
this  is not visible on the scale of Fig. \ref{fig2} but can be
seen in Fig. \ref{fig2sub}.
As long as $W_0^{{(out)}}$ spirals around FP0, it has to intersect
$W_1^{(in)}$, which means that there is a hetero-clinic
orbit and therefore a stationary domain-wall.
When $c$ approaches $c_{crit}$, this spiraling motion becomes less 
prominent,
and consequently, the intersection of $W_0^{{(out)}}$ and $W_1^{(in)}$
shifts to FP0 (see Fig. \ref{fig2sub}). As a consequence,
the position of the domain-wall shifts to large values of $x$.
 
When $c$ has crossed the critical value $c_{crit}$,
FP0 is a saddle and it turns out that  $W_0^{{(out)}}$ and $W_1^{(in)}$
only intersect in FP0 itself (see Figure \ref{fig2}d);
this corresponds to a domain-wall that is shifted to infinity.

\subsubsection{Validity of $c_{crit}$.}
There are two different assumptions involved in the calculation
of $c_{crit}$. First of all, $\gamma$ needs to be small, in order
for the perturbation equation (\ref{balance}) to hold.
Secondly, we used for $\gamma(x)$ a step-function.
We will now  discuss the validity of
$c_{crit}$ when $\gamma(x)$ is of more general form.
 
The latter assumption is, due to
the scaling properties of equation (\ref{balance}),
 not crucial, as we have indicated above.
When we rescale $\gamma\rightarrow \delta \gamma$, with
$\delta<1$, we can compensate for this by rescaling
the spatial coordinate; effectively, the function $\gamma(x)$ becomes
 steeper then.
When
$\delta \downarrow 0$, $\gamma(x)$ becomes infinitely small and
infinitely steep. Therefore, the results that are obtained
when $\gamma$ is a step-function, are still valid when $\gamma(x)$
is an arbitrary monotonically decreasing function, provided that $|\gamma|
$ is small.

It is tempting to try to extend the phase-space analysis to the 
non-perturbative case, i.e., when $\gamma$ and $c$ are not
small. The stationary version of the coupled NLD equations
(\ref{FK}) can then be written as 
a 4-dimensional dynamical system. The 
equations for the trajectories cannot be solved in this case, but
we can inspect the eigenvalues of the fixed points that
correspond to the single-mode and bimodal states. The latter fixed point
always has two positive and two negative eigenvalues. 
The fixed point corresponding to the single-mode state changed from
a saddle to a spiral at $c=c_{{crit}}$ in the perturbative case,
and something similar happens in the 4-dimensional system. 
The eigenvalues of this fixed point are given by the expressions
$\frac{1}{2} (c \pm \sqrt{c^2 + 8})$ and
$\frac{1}{2} (c \pm \sqrt{c^2 + 4 \gamma_{{ min}}})$.
One can readily find that, when $c=-2 \sqrt{-\gamma_{{ min}}}$, the
fixed points changes from being a saddle (two negative
and two positive eigenvalues) to a fixed point
with one positive, one negative and two complex conjugated eigenvalues.
Although this does not prove that the stationary DW 
disappears, this is at least
an indication that the previously obtained value $c_{{ crit}}$ may be
also valid when $\gamma$ and $c$ are not small.

\subsection{Numerical simulations}\label{secFKnum}
 
To test the validity of the perturbation equation (\ref{balance})
and to verify our predictions, we have performed numerical simulations
of the coupled NLD equations (\ref{FK}) 
in a periodic system of size $L=400$.
Because of the periodicity of the system,
there are two  domain-walls but we will, as before,
focus attention on the domain-wall that connects a
single-mode state $(B=0)$ to the left with a bimodal state
to the right.
The results reported here were obtained by using a pseudo-spectral
code with the time step 0.05 and, typically, 256 modes; runs with a higher
number of modes up to 1024 were performed to check the results. 
The code was such that $A$ and $B$ could either be real or complex-valued
in order to simplify the comparison between the results for 
the NLD equations and the RGL and CGL equations later on.

When $c \downarrow c_{crit}$, the intersection point of the
manifolds approaches FP0, and consequently the position of the
DW shifts to $+\infty$. In this limiting case, a single-mode state 
intrudes into the domain where
$g<1$, where this state is unstable. Therefore, one might expect
that the DW is unstable when $c$ is still slightly above 
$c_{{ crit}}$. 
 We find below that this is the case indeed, and
we will denote the value of the group-velocity
where the domain-wall turns unstable by $c_i$.

The simulations carried out below focus on three items. 
First of all, we will
verify that the stationary DW predicted by the phase-space
analysis exists and is stable when $c$ is not too close to
$c_{{ crit}}$. 
Subsequently, we determine, for various
choices of the inhomogeneity, the value of $c$ where the domain-wall
turns unstable, and compare it with $c_{crit}$.
Finally, we investigate the dynamical states that occur when
$c$ is decreased beyond $c_i$.

\subsubsection{Stationary states}
In this section we present the results of the simulations
of the NLD equations for $|c|<|c_{{ crit}}|$
and step-like $\gamma$.
Most of the phase-space analysis presented above was based on the
assumption that $\gamma$ and $c$ were small
and we therefore take
$\gamma(x) = 0.1$ at $x<0$, and $-0.1$ at $x>0$; in this case,
the critical value of the group-velocity is
$c_{crit}=-2\sqrt{-\gamma_{min}}\approx -0.632$.

Because of the periodicity of the system,
there is also a step of $\gamma$ and, hence, another DW
at $x=\pm 200$, but we will, as before,  
focus attention on the DW solution near $x=0$ that connects a
single-mode state $(B=0)$ to the left with a bimodal state
to the right.

We have found that for $|c|\lesssim 0.6$
the system relaxes to one of the stationary DW's that are shown 
in Fig. \ref{statfig}. 
These domain-walls were obtained by numerical
simulations with $1024$ modes.
These numerical results clearly demonstrate that the
stationary DW's are stable when $c$ is not very close to 
$c_{{ crit}}$; this fact cannot be derived from the
phase-space analysis alone.

Comparing the domain-walls obtained  from direct
simulations of the NLD equations with those obtained by numerical
integration of the ordinary differential equation
$d\chi/dx =\psi$ along the analytically obtained flow-lines,
we found that the shape of the domain-wall is predicted very
well for all values of $c$ that we consider here.
The largest deviations
occur for $|c|=0.6$, when the DW's obtained by the two
aforementioned methods have a relative spatial shift $\approx 2$.
The quantity $R^2$ (see Eq. \ref{chi})) is equal to $1\pm0.005$ for the single-mode,
and $0.95 \pm 0.005$ for the bimodal state (note that from
$|A|=|B|=\sqrt{1/(1+g)}$ it follows that $R^2 =0.95$ in the
bimodal state).

\subsubsection{Instability of the domain-wall}
 
Now that we have checked that the predications that follow from
the perturbation theory are correct  when $c$ is not too close
to $c_{crit}$, we will investigate what happens when $c \downarrow 
c_{crit}$.
The numerical simulations that we will present below reveal that
the strength and the
steepness of the inhomogeneity, the value of $c$ and
the initial conditions all may influence the dynamics.
To sample the parameter space without going into too much detail, we have
eventually restricted ourselves to inhomogeneities of the form
\begin{equation}\label{inhomoeq1}
g(x)= 1 + \Delta g \tanh(s(x-0.25L))\tanh(s(x+0.25 L))~,
\end{equation}
where $s$ stands for the steepness of the inhomogeneity and $L$
is the size of the system.
The middle part of the system is where the bimodal state exists.
We have shifted the inhomogeneities away
from $x=0$ because this makes the pictures for
the time evolution of $A$ and $B$ that are presented below more clear.
 
We take values of the inhomogeneity strength
$\Delta g$ of $0.05,0.1,0.2$ and $0.3$ in order to obtain information
on the dependence of the dynamics on the strength of the inhomogeneity.
For the steepness $s$, two different values were selected:
$0.1$ for a smooth inhomogeneity, and $10$ for a steep inhomogeneity;
for the numerical simulation presented below, this last value
is practically equivalent to a step-like inhomogeneity.
 
As can be seen in Fig. \ref{statfig}, the location of the domain-wall 
shifts to larger values of $x$ when
$c$ approaches $c_{crit}$,
in agreement with the phase-space analysis.
In this case, a
large patch of the unstable single-mode state intrudes in the $g<1$ 
domain.
Eventually,
this renders the domain-wall state unstable, and we define $c_i$
as the group-velocity for which this instability occurs
(see Fig. \ref{schemafig1}).
 
We found that the primary instability indeed occurs when $c$ has 
approached
$c_{crit}$ within a few percent.
The large patch of  unstable single-mode state turns then
convectively unstable, but the total domain-wall state is 
absolutely stable.
This means that noise, which is mainly due to the
discretization of space and time in the numerics, is amplified
in the region where we have the unstable single-mode state.
Due to the group-velocity, these fluctuations are advected towards
the  stable bimodal state (mode $A$) or the
stable single-mode state (mode $B$), where they are dissipated.
To monitor where this instability occurs, we have followed the time
evolution of the norm $\int dx A$ as a function of the group-velocity.
When the  instability of the domain-wall state  occurs,
the norm starts to oscillate with well-defined frequency.
We checked for hysteric effects but could find none,
and we conclude that the
domain-wall turns unstable via a forward Hopf-bifurcation.
However, the fact that in this case the system is still absolutely stable,
yields that hidden line plots of the domain-wall
hardly show a clear dynamic state.  When $c$ is close
the $c_i$, one can observe the instability best in the norm 
$\int dx A$.
 
We have measured $c_i$ for $s=10$ and various values of $\Delta g$, and
the results are listed below.
\begin{center}
\begin{tabular}
{|c|c|c|c|} \hline
$\Delta g = |\gamma_{min}|$ & $c_i$ & $c_{crit}$ & $c_i/c_{crit}$\\  
\hline
0.05 & -0.43 & -0.447 & 0.96 \\ \hline
0.1 & -0.62 &-0.632 & 0.98\\ \hline
0.2& -0.87 & -0.894 &0.97 \\\hline
0.3 &-1.06 &-1.095 & 0.97 \\\hline
\end{tabular}
\end{center}
The error-bar on the measurements of $c_i$ is of order $0.01$.
The most important conclusion that we can draw from this measurements 
is that
$c_i$ is very close to $c_{crit}$, and that this is independent of
the value of $\gamma$.
 
\subsubsection{Dynamical states}

In this section we investigate the fate of the domain-wall when
$c$ is decreased beyond $c_i$.
It is important
to emphasize that, unlike many other transitions to
dynamical behavior, there is, according to our phase-space
analysis, no stationary albeit unstable state
when $|c|>|c_{crit}|$. Thus, the single-mode and bimodal states
are stable, each in its own domain, but there is no
{\em stationary} domain-wall to connect these two asymptotic states.

We study again inhomogeneities of the form (\ref{inhomoeq1}),
and take the group-velocity
$1.1, 1.2$, $1.5$ and $2$ times the critical value $v_{crit}$.
As initial conditions, we always took
\begin{eqnarray}\label{initialcon}
A(x) &=& 0.85 + 0.15 \tanh((x-0.25 L)/10) \tanh((x+0.15 L)/10), \\
A(x) &=& 0.35 - 0.35 \tanh((x-0.25 L)/10) \tanh((x+0.15 L)/10), 
\end{eqnarray}
which is close to the stationary state for $c=0.9 c_{{ crit}}$.

Most of the dynamical states fall into two distinct
types of behavior. For the weak inhomogeneities, the DW moves irregularly 
back and forth around an average position. 
For stronger inhomogeneities, we often found that traveling kinks 
connecting states with the
opposite signs of the amplitude $B$ were generated in the bimodal regime. 

We will describe in detail the dynamics that are found
for the inhomogeneity with the steepness $s=10$, which is the value
of $s$ that we will focus on; the other value of $s$ will merely serve
to explore the generality of the behavior found for $s=10$.

At $\Delta g =0.05$, we observe a disorderedly moving DW, and this is also
the case for $\Delta g =0.1$. 
We will focus now on $c=1.1 c_{{ crit}}$ and $\Delta g =0.1$.
The dynamics of the amplitudes $A$ and $B$ and the quantity $N:=
\int_{-\infty}^{+\infty}dx A/<\int_{-\infty}^{+\infty}dx A>$ are shown in 
Fig. \ref{figfkv1}. As can be clearly seen in this figure, an
essential dynamical degree of freedom is the position of the DW
that moves irregularly around a certain mean position.
This mean position shifts to the right when $c$ decreases.
By extending the simulations to longer time intervals
we have checked that the motion remains disordered.
 
To inspect whether the dynamics are chaotic, we have
studied effects of small perturbations in the initial conditions.
We have performed two runs, one (the unperturbed run) starting from
initial conditions obtained as the final
state of previous simulations in the disordered regime;
for the perturbed run, the initial profile  of $A(x)$ was unaltered, but 
the value of $B$ at $x=0$ was diminished by $10^{-5}$.
The data pertaining to the perturbed run will be distinguished by a prime.
We have plotted $|A-A'|$ in Fig. \ref{fkv2}b, and $N$ and
$N':= \int_{-\infty}^{+\infty}A'dx/<\int_{-\infty}^{+\infty}
dx A>$ in Fig. \ref{fkv2}a. Clearly, the
dynamics are sensitive to the small perturbation.
The perturbation of the initial data is seen to grow and spread out,
but only an area around the DW is affected (Fig \ref{fkv2}b).
When a perturbation far from the DW is initiated,
the same effect is eventually observed, but after a longer
transient time, which is presumably the time the perturbation
needs to reach the region around the DW.
For the motion to be chaotic, we need to find
an {\em exponential } divergence between $A$ and $A'$ for
some time-interval. A close inspection of the dynamics
for $A$ and $A'$ reveals that this is {\em not} the case;
$A$ and $A'$ diverge but not exponentially.
Therefore, we suspect the disordered motion of the domain-wall
to be due to the convectively amplified discretization noise.
This scenario is somewhat similar to the
dynamics described in \cite{chomaz}.

It seems that the dynamics, at least at the
lowest order, can be described by a disordered
drift of the DW, which leaves
its shape intact. To check this,
we have collapsed 100 snapshots from the time evolution of A,
plotting $A$ vs. $dA/dx$. All the curves fall, with a good approximation,
on top of each other, which indicates that, to the lowest order,
the dynamics can be adequately modeled by the only degree of freedom, 
viz., the position of the DW. 

When the strength of the inhomogeneity is increased, 
a different type of behavior is observed. As an example, we will
consider what happens at $\Delta g =0.2$, although qualitatively
the same behavior occurs at larger $\Delta g$. 
The motion that occurs at $c=1.1 c_{{ crit}}$ is shown in Fig. \ref{perdyn}.
A close inspection of the data reveals that, to the left of the DW,
where $B$ is very small, zeroes of the function $B(x)$ are generated 
periodically. Each of them is then convected and
amplified, leading to the generation of a traveling kink.
We have found that the frequency of the kink generation goes
approximately linearly
with $c$, and that it is nonzero at $c=c_{{ crit}}$. 
The velocity of the kinks is slightly smaller than the group-velocity.
This is not surprising: associated with the kink in
$B$ is a small ``bump'' in $A$ (see Fig \ref{fkv2}). Without this bump,
the kink would travel with precisely the group-velocity of $B$, 
but since the group-velocity of $A$ is opposite to that of $B$, the
small bump somewhat lowers the velocity of the kink.

Taking a smoother inhomogeneity ($s=0.1$) suppresses the generation of
kinks, and if they are still generated, this process is mixed with the irregular
motion of the DW, as is shown in Fig \ref{mix}. 
The generation of kinks can even become irregular.

In conclusion, when $c$ is beyond $c_{{ crit}}$, the dynamical state
may be represented by the disordered DW, regularly generated and
traveling kinks, or a mixture between the two.

\section{The RGL equations with the group-velocity}\label{rglsec}

In this section we relax the condition that
the order parameters $A$ and $B$ are real,
and study the behavior of a DW
in a system of two coupled RGL equations with the group-velocity terms:
\begin{mathletters}\label{RGL}
\begin{eqnarray}\label{rglsys}
\partial_t A  + c \partial_x A &=& A + \partial_x^2 A -
(|A|^2 + (1 + \gamma(x)) |B|^2) A ~,\\
\partial_t B  - c \partial_x B &=& B + \partial_x^2 B -
(|B|^2 + (1 + \gamma(x)) |A|^2) B ~.
\end{eqnarray}
\end{mathletters}
The main question here is to see whether the results obtained
for the NLD equations  are relevant for the RGL equations.

The complex order parameter poses the question of wavenumber selection,
and the possibility of nonzero wavenumbers constitutes the main
difference with the NLD equations. 
We have found that, for the stationary states, the wavenumbers
often, but not always, relax to zero when the initial conditions
have nonzero wavenumbers. 
 When the wavenumbers of $A$ and $B$
are constant and equal, the NLD equations
can be used to describe the stationary states;
when the wavenumbers are more general, modified NLD equations
model the stationary domain-walls, as we will discuss below.
In the dynamical regime, small wavenumbers may be generated starting from
initial conditions with zero wavenumbers.

Note that the usual Eckhaus band of stable wavenumbers for a single RGL
equation 
\cite{Eckhaus}  will be affected
by the coupling between the two RGL equations. In particular, since for
$\gamma\rightarrow 0$ both the single-modes  and the
bimodal states with zero wavenumber are only marginally
stable, one may expect that the size of the band of stable wavenumbers
is directly related to $\gamma$; and that for $\gamma\rightarrow 0$
the band of stable wavenumbers closes.
We have not pursued this question further.

The wavenumbers of $A$ and $B$, which we refer to
as $q_A$ and $q_B$, also have an influence on the stability borders of
the single and bimodal states. When $q_A=q_B$, the crossover between
single and bimodal states occurs at $g=1$, just as when
$q_A=q_B=0$. This can be shown
by substituting plane-wave solutions with equal wavenumbers in the RGL
equations 
and performing a rescaling of $x$, $t$ and the amplitudes, similar to
the rescaling
that is used to scale out the growth-rate $\varepsilon$. 
Apart from this scale transformation,
the equations for zero wavenumber and equal wavenumber are equivalent.
When $q_A \neq q_B$, the stability borders for both
the homogeneous single-mode and bimodal states
shift away from $g=1$, and in particular there exists
a tiny parameter regime  $g'<g<g''$,
for which both the  single-mode and bimodal states
are linearly unstable. The differences between
of $g'$ and $g''$ and $1$ are of order
$q^2$, and we will not go into this in detail.

\setcounter{subsubsection}{0}
\subsection{Stationary states}\label{rglstatsec}
 
In this section we focus on the case $|c|<|c_{{ crit}}|$ and
explore  numerically the mechanisms by which the wavenumber relaxes.
The values of $c_i$ that we found are
very close to those found in the simulations
of the NLD equations (we will come back to this below).

The simulations of Eqs. (\ref{RGL}) were performed for
a periodic system of the size 400. 
The inhomogeneity was of the form given in
Eq. (\ref{inhomoeq1}) with $s=10$. 
The initial conditions were chosen to be DW-like, similar to 
Eq. (\ref{initialcon}), but the wavenumbers $q_A$ and $q_B$ of the complex
variables $A(x)$ and $B(x)$ in
the initial states were allowed to be nonzero but constant,
to study the wavenumber relaxation.

In the variational case $(c=0)$ the system (\ref{rglsys}) 
has a Lyaponav functional
${\cal L}$ with a minimum at $q_A=q_B=0$. 
However, such states cannot always be reached, since
the periodic boundary conditions lead to
conservation of the total phase difference across the system,
as long as the amplitude remains nonzero.
So, the wavenumber of the, e.g., mode $A(x)$ can be relaxed through the 
so-called phase-slips occuring at the points where $A=0$ vanishes 
\cite{phaseslip}. When $A$ is not close to zero, generation of such a
phase-slip may lead to a significant increase of
${\cal L}$, which is forbidden. It is then possible for the
system to end up in a {\em local} instead of a {\em global}
minimum of ${\cal L}$. But when $A$ is small in some region,
phase-slips easily occur, and the wavenumber relaxes to zero.

When $c\neq0$, the RGL equations are no longer variational,
but for the stationary domain-walls the tendency to evolve to states
with zero wavenumber is still present.
We found that, the wavenumber of the $B$ mode
always relaxes to zero,
while for the $A$ mode, the wavenumber occasionally remains nonzero.

The $B$ mode is close to zero in the single-mode region. Phase slips
can easily occur in this regime, and we have observed that
$q_B$ always relaxes to zero,
both for $v_g=0$ and $v_g \neq0$.
At $c=0$, the local wavenumber diffuses, and, at $c\neq0$,  the wavenumber 
is advected to the single-mode region, where it relaxes.
When the initial $q_B$ is large, the phase slips of $B$
occasionally occur in the double-mode region, but only
as a transient.

For the $A$ mode, two different mechanisms for relaxation
of $q_A$ are observed.
In the simplest case, the phase slips occur close to the DW, until $q_A$ 
falls to a level $\sim 0.1$.
It may also occur that the phase slips lead to a single-mode state
consisting of alternating  patches with $A=0$ and $B=0$.
In that case, both $q_A$ and $q_B$ relax to zero. 
The kink that is generated between these two different single-mode
states is not a global, but rather a local
 minimum of ${\cal L}$ corresponding to  a metastable state. 

The stationary domain-walls of the coupled RGL equations are very similar
to those of the NLD equations.
When $q_A=q_B$, we can show this by substituting plane-waves
of the form $A=|A|\exp(iqx)$ and $B=|B|\exp(iqx)$ into the RGL equations,
which yields that the RGL equations are similar to the NLD equations,
up to the coefficient $(1-q^2)$ in front of the linear term.
By the aforementioned scaling we can scale this factors out,
and this yields the NLD equations for $|A|$ and $|B|$.
Therefore, when $q_A=q_B$, the stationary domain-walls produced by the 
RGL equations
can be described by the NLD equations.
The critical value of the group-velocity where the position
of the domain-wall diverges is then also precisely given by
the value of $c_{crit}$ obtained for the NLD equations. The effect of
nonzero wavenumber  on
$c_i$ can in principle be more subtle, but in practice
we could not find a difference between the values of $c_i$ for
the NLD equations and the RGL equations with $q_A=q_B$.
 
When $q_A\neq q_B$, we cannot scale out the wavenumbers.
Substituting plane-waves
of the form $A=|A|\exp(iq_Ax)$ and $B=|B|\exp(iq_Bx)$ into the RGL
equations,
yields two NLD-like equations were the first terms of the right-hand
side $A$ and $B$ are replaced by $(1-q_A^2)A$ and $(1-q_B^2)B$.
Only one of these pre-factors can be scaled out, so the
ordinary NLD equations are not correct. However,
since the difference of $q_A$ and $q_B$ that occurs in
stationary domain-walls is of order $0.1$ at most, and the
effect on the RGL equations is of order $q^2$, the NLD equations
are still valuable to give a lowest order description of the ensuing
domain-walls.
In principle, one could carry out the geometrical analysis
for the equations with the $(1-q_A^2)A$ and $(1-q_B^2)B$
terms; we will not give all the details here, but it can be shown
that $c_{crit}$ is perturbed by terms of order $q^2$.
This is consistent with the numerically observed
increase of the value of $c_i$
with $q_A$ (when $q_B=0$); this value increases at most
a few percent when $q_A =0.1$.

\subsection{Dynamical states}

To investigate the dynamical solutions of the RGL equations, we
used the same inhomogeneities and values of $c$ 
as for the NLD equations. The same initial conditions
were used, so we restrict ourselves to the zero initial wavenumbers.
For a steepness $s=10$, we have found for almost all $c$ 
beyond $c_{{ crit}}$ and $\Delta g$ a
disordered fluctuating DW state. The mean position of the
DW shifts to the right with decreasing $c$; this effect becomes
weaker when $\Delta g$ becomes larger.
A slightly different state was observed for 
$\Delta g =0.1$ and $c=2 c_{{ crit}}$, when the oscillations
of the DW showed a strong periodic component but are nevertheless 
disordered; the reason for this  is not clear.
We found that when $\Delta g$ is increased, 
the fluctuations of the
position of the domain-wall in general decrease.
 
The phase of $A$ remains zero in the dynamical state. The phase
of $B$ departs from zero, and $B$ slowly develops wavenumbers $\sim
0.01$. At longer times, these wavenumbers often remain constant in
space and time. A clear exception is the
aforementioned state with $\Delta g =0.1$ and $c=2 c_{{ crit}}$, where
periodic modulations of the wavenumber persist.

For smoother inhomogeneities (s=0.1) and small $\Delta g$, 
the wave number generation is suppressed, and a traveling
kink state is observed.
This state is similar to the states produced by the NLD equations, in the 
sense that the phase difference across such a kink is exactly $\pi$. 
However, the generation of kinks is now no longer periodic.
When $\Delta g > 0.1$, these kinks are not produced,
and the local wavenumber is generated instead, just as in the case of the
steeper inhomogeneity.

In conclusion, both the stationary states and the
critical value of $c$ for the RGL equations are almost the same as
for the NLD equations, while the dynamical states are somewhat different. 
The main mechanisms
that play a role for the domain-walls of the RGL equations can be
described by the NLD equations.

\section{The CGL equations}\label{cglsec}

In this section we will study the coupled CGL equations:
\begin{mathletters}\label{cgl}
\begin{eqnarray} 
\partial_t A  + c \partial_x A &=& A + 
(1+ i c_1) \partial_x^2 A -
\left[(1-i c_3) |A|^2 + (1-i c_2) g(x) |B|^2\right] A ~,\label{cgleqa}\\
\partial_t B  - c \partial_x B &=& B + 
(1+ i c_1) \partial_x^2 B -
\left[(1-i c_3) |B|^2 + (1-i c_2) g(x) |A|^2\right] B ~. \label{cgleqb}
\end{eqnarray}
\end{mathletters}
These are the generic amplitude equations for left- and right traveling
waves, and the group-velocity terms appear here naturally.
The coefficients $c_1,c_2$ and $c_3$ can be obtained from a systematic
expansion of the underlying equations of motion and play an essential role
for the dynamics of the CGL equations. The nonlinear dispersion
coefficients $c_2$ and $c_3$ are, in general, not equal, and we will
find below that their difference will be a crucial parameter for the
DW's.
The behavior of a single CGLE is already incredibly rich \cite{CH},
and the situation for the coupled CGLE's is of course not simpler.
As a function of the coefficients $c_i$, typical states in homogeneous
coupled CGLE's
include single and bimodal phase-winding solutions,
periodic solutions and spatio-temporal chaotic solutions 
\cite{Coullet,saka}.
For fixed values of the coefficients, different states can coexist.
In the following we will restrict ourselves to describing some
of the interesting states that occur in the numerical simulations
of the coupled CGLE's when there is a CC
that passes through its critical value.

The numerical simulations were carried out similarly to the RGL model,
in a periodic system of the size 400.
The inhomogeneity was of the form given in Eq. (\ref{inhomoeq1}),
and we focus on the case $s=0.1$, although we have performed some runs with
$s=10$. It turns out that only the details of the dynamics are different
for this larger value of $s$.
As initial conditions we
use, as before, the stationary state that is obtained when $c=c_1=c_2=c_3=0$.
This means in particular that the critical value for $g$ is
$1$; it has been shown by Sakaguchi \cite{saka}, that for periodic
or disordered states the transition between single and bimodal behavior
can occur for values different from $1$.

Even with these restrictions, the parameter-space of the
CGLE's is too large to warrant a complete overview of the dynamical
behavior.
The nonlinear dispersion is essential for the fate of the domain-wall, and
to restrict the search in parameter-space, we restrict ourselves to
three different cases: (A) $c_2=c_3$,
(B) $c_2=-c_3$, (C) $c_2=0$.
The role of the group-velocity and $c_1$ is discussed briefly
for each case.

We have found many interesting dynamical states;
in many cases, the homogeneous single and bimodal states appear 
to be invaded
by more complicated states that grow from the initial DW.
By carefully adjusting the various parameters,
it seems possible to smoothly proceed from regular periodic
to completely disordered states. Below we will only sketch a few of the
dynamical regimes that we have found. A more systematic exploration
is left for a further work. It is relevant to note that, when 
the CC constant
is far from its critical value 1, we did not find new behavior.

\subsection{The case $c_2=c_3$}

When the nonlinear dispersion coefficients $c_2$ and $c_3$ are equal,
the nonlinear term of Eq. (\ref{cgleqa}) (similar for (\ref{cgleqb}))
can be written as $(1-i c_2)(|A|^2 + g(x) |B|^2) A$.
Just as in the RGL case, ($|A|^2 + g(x) |B|^2$) differs only slightly 
from the value 1
for a DW state, so the nonlinear dispersion
acts similar to the linear dispersion.
The main effect of the dispersion is then
to shift the wavenumber of the $B$-mode, 
but apart from this, the behavior of the CGL's is 
qualitatively similar to that of the coupled RGL's.  

When $q_B\neq 0$, $|B|$ is smaller then $|A|$ in the bimodal regime,
and as a result, the critical group-velocity is seen to increase
(see the discussion
on the effect of nonzero wavenumber on $v_i$ for the RGL equations).
For example, when $c_2=c_3 =0$ and $c_1 = \pm 1$, 
the selected wavenumber of $B$ is close to $0.1$,
and $|B|$ is a few percent smaller than $|A|$.
The critical value of the group-velocity is then
increased by a few percent.
When both $c_1$ and $c_2 = c_3$ are different from zero, the wavenumber
of $B$ can be large enough to shift the critical group-velocity substantially.
For $c_1 = -1$ and $c_2=c_3=0.5$, the wavenumber of $B$
is close to $0.2$, and the amplitude of $B$ is approximately
20 percent smaller than
that of $A$. The critical group-velocity is then found to be
between $c=-0.84$ and $c=-0.85$. Although this is quite different
from the case where the wavenumbers are zero,
the essence of the transition to the dynamical DW's is still given by
the analysis for the NLD equations.
When the difference between $c_2$ and $c_3$ is small, we 
find qualitatively the same behavior
as for $c_2=c_3$.

\subsection{The case $c_2\neq c_3$}

When $c_2 \neq c_3$, the nonlinear dispersion is no longer
spatially independent, and this leads to oscillatory, 
spatially periodic or chaotic behavior.
When we take both the linear dispersion and the group-velocity equal 
to zero,
we already find various types of behavior, as shown in Fig. \ref{cglfig1}.
The oscillatory behavior shown in Fig. \ref{cglfig1}a and in
more detail in Fig. \ref{cglfig2} arises from a feedback
mechanism between local wavenumbers and amplitudes. When
$c_2\neq c_3$, gradients of $|A|$ and $|B|$ generate local
wavenumber (see Fig. \ref{cglfig2}b), and
this local wavenumber suppresses, via the diffusive term,
the amplitudes.  As a result of 
this feedback mechanism, the DW becomes oscillatory. 

The strength of
the aforementioned feedback mechanism grows with the 
nonlinear dispersion, and above a certain threshold, we find that
the oscillations do not stay confined around the DW, but spread
out into the single-mode region. For instance, for $c_3=-c_2 =0.2$ we
found that a periodic state is generated (Fig. \ref{cglfig1}b), 
that is similar to the periodic state described recently 
by Sakaguchi \cite{saka}. The DW itself becomes disordered.
It should be noted that for these values of the coefficients,
there are
 asymptotic single and bimodal phase-winding solutions that are linearly 
stable.

Finally, the nonlinear dispersion can become so large that
also the bimodal state becomes periodic. The disorder seen
in Fig. \ref{cglfig1}c may be either a transient behavior or 
an established state; our simulations were not conclusive,
and we leave this for further work.
Here the coefficients of the
CGL equations 
are such, that homogenous phase-winding solutions are unstable,
and that even in homogeneous systems, i.e., for $\gamma$ fixed at
$\pm 0.1$, periodic states arise. The domain-wall here is disordered,
and so this state shows the competition between linearly stable
periodic states and a disordered domain-wall.

The linear dispersion has a damping effect on the dynamics as
for $c_1\neq0$ similar behavior is observed
for slightly higher values of the nonlinear dispersion.
A nonzero group-velocity has a more complicated effect on the dynamics,
as it  breaks the reflection symmetry.
The oscillations that occur for small nonlinear dispersion are damped,
because the group-velocity terms advect the local
wavenumbers away from the DW, and therefore they suppress the 
aforementioned feedback effect.
For instance, for
$c_1=-1$ and $c=0.5$, we have observed stationary states up to
$c_3=-c_2 =0.1$.
On the other hand, the symmetry breaking can have a destabilizing effect 
on the dynamics:
the periodic state shown in Fig. \ref{cglfig1}b becomes disordered
when $c$ is nonzero. 

When we move away from the line $c_2=-c_3$, we find that
spatio-temporal chaos occurs quite easily. We have focused on the 
case $c_1=c_2=0$.
For $c_3=0.2$, a periodic state similar to the one depicted in
Figure \ref{cglfig1}b occurs. When $c_3$ is increased, both the periodic
and the bimodal states become gradually disordered 
(Figs. \ref{cglfig3}a and \ref{cglfig3}b).
 
When $c=0$, the chaotic state consists of more or less stationary, irregularly
growing and decaying pulses, but when $c\neq0$, this quasi-stationary
character is destroyed. We have checked that in the disordered
regime, two slightly different initial conditions diverge throughout
the whole domain, which shows that these states are an example
of spatio-temporal chaos.
Note that this occurs for values of $c_1,c_2$ and $c_3$ for which
the homogeneous single and bimodal state are linearly stable.
Apparently the DW acts as a ``seed'' for the disorder, that then
spreads out and completely destroys the plain-wave state.

We have concentrated here on the case $c_2=-c_3$, although it should be
stressed that when $c_2$ is not exactly $-c_3$, we observed similar 
behavior.
The choice $c_2=-c_3$ merely serves to limit ourselves in exploring the
parameter-space.

\section{Conclusions}

In this paper we have considered domain walls
between single-mode and bimodal states for three types of
coupled equations with a spatially dependent coupling coefficient. 
In the simplest case of two coupled NLD equations with the
group-velocity terms, we were able to
reduce the description of stationary configurations to a single 
non-autonomous second-order
ordinary differential equation, that was used to
determine analytically a necessary condition for the existence of a
stationary DW in terms of the group-velocity. We have found
that our prediction for the destabilization of such a stationary
DW is in good agreement with numerical simulations, and we have found
chaotically oscillating DW's in the case when the group-velocity is
beyond the corresponding threshold. 
For two coupled RGL equations we have found a similar scenario. Finally, for
the coupled CGL equations, we have found that, in most cases,
the DW's are unstable, even when the group-velocity is zero, and 
spatio-temporal disordered states often occur in this model.

In the future, it would be interesting to investigate the
competition between the various states of the coupled CGL equations.
In particular, not much is known about the periodic states
that seem to play an important role here. Possible
research subjects include the development of analytical solutions,
the development of counting arguments \cite {saarloos} 
and the competition between periodic and phase-winding solutions.
The effect of an inhomogeneity as studied in this chapter
on the various states may be a valuable tool in probing the
states that occur for constant $g$.
The effect of nonzero group-velocity on the domain-walls in
the CGL equations is poorly understood; it would be interesting to
see whether  the divergence and subsequent instability of the
domain-wall, as observed for the NLD equations and CGL equations,
still has some relevance for the CGL equations when $c_2$ and $c_3$ are
sufficiently different.

\section{Acknowledgment}
 
We appreciate useful discussions with Wim van Saarloos, Willem van de Water
 and Bert Peletier.

\newpage

\begin{figure}
\caption[]{The exact DW solution (\ref{exact}) to the reduced NLD equations
without the group-velocity terms, when the inhomogeneity is chosen as
$\gamma(x)= -\kappa^2 \tanh (\kappa x)$, with $\kappa =0.2$.\label{fig1}}
\end{figure}
\begin{figure}
\caption[]{Trajectories of the two-dimensional dynamical system 
(\ref{dynamical}) for the variables $\chi$ and
$\psi$ in the case when $\gamma(x)$ is the step function (\ref{step}) 
with $\gamma_{{ min}}=-0.1$ and $\gamma_{{ max}}=0.1$. 
The fat dots represent the fixed points FP0 and FP1,
while the bold and dashed curves represent, respectively, their unstable and
stable manifolds $W_0^{{ out}}$ and $W_1^{{ in}}$.\label{fig2}}
\end{figure}
 
\begin{figure}[tbp]
\caption{The behavior of  $W_0^{{(out)}}$ (bold curve) and $W_1^{{(in)}}$
(dashed curve)
around FP0 (black dot) for a range of values for the group-velocity.
In this case we have taken $\gamma_{min} = -0.1$, which yields a critical
velocity $c_{crit} \approx 0.632$.}
\label{fig2sub}
\end{figure}

\begin{figure}
\caption[]{The function $\chi(x)$ corresponding to the domain-wall
solutions for various choices of the group-velocity
$c$. The domain walls were obtained by direct simulations
of the coupled NLD equations (\ref{FK}). The dashed curve corresponds
to the DW for $c=0.6$, while the fat curve corresponds to
$c=-0.6$. The DW's in between correspond to $c=0.4$, 0.2, 0, -0.2 and -0.4,
respectively; the DW shifts to
the right when $c$ is decreased towards $c_{{ crit}}$. \label{statfig}}
\end{figure}
 
\begin{figure}[tbp]
\caption{Schematic representation of the bifurcation
structure as a function of the group-velocity for weak inhomogeneities.
The vertical axis symbolizes the position of
the domain-wall.
When $c$ is decreased (this corresponds
to moving to the right on the vertical axis)
towards values close to $c_{crit}$,
the position of the domain-wall  diverges (bold curve).
For $c= c_i$, the domain-wall turns unstable via the
occurrence of a Hopf bifurcation. When $c$ is
decreased even further, various dynamical states occur (dotted curve).
For $|c_i|<|c|<|c_{crit}|$, there is, according
to the phase-space analysis, a stationary, but
unstable domain-wall (dashed curve); for
$c$ beyond $c_{crit}$, there is {\em no}
stationary domain-wall.} \label{schemafig1}
\end{figure} 
\begin{figure}

\caption[]{The upper panel shows the chaotic fluctuations of
$\int_{-\infty}^{+\infty} dx A / <\int_{-\infty}^{+\infty}
dx A>$, where the average is over the entire time
of the simulation. The plots of $A$ and $B$ for this 
chaotic state are shown in the lower panels. 
Subsequent snapshots have a time difference of 80 and
are shifted in the upward direction by a distance of 0.05.
We used an inhomogeneity of the form (\ref{inhomoeq1}) with $s=10$
and $\Delta g= 0.1$, and set the group-velocity $c$ to $1.1 
c_{{ crit}}$.
The left DW is seen to fluctuate chaotically
whereas the right DW is completely stationary. \label{figfkv1}}
\end{figure}

\begin{figure}
\caption[]{The sensitivity of the DW to a perturbation of the initial 
conditions is demonstrated using evolution of two close initial conditions
(see the text). 
The upper panel shows the divergence of
the time evolution of $\int_{-\infty}^{+\infty} dx A$
(the thin curve) and $\int_{-\infty}^{+\infty}dx A^{\prime}$ (the
fat curve) divided by $<\int_{-\infty}^{+\infty} dx A>$. The lower panel 
shows plots of the difference between $A$ and $A^{\prime}$; consecutive
snapshots have a time difference of 40 and
are shifted in the upward direction by a distance of 0.001. \label{fkv2}}
\end{figure}

\begin{figure}
\caption[]{
The upper panel shows the periodic fluctuations of
$\int_{-\infty}^{+\infty}dx A / <\int_{-\infty}^{+\infty}dx A>$. The
plots of $A$ and $B$ for this periodic traveling kink state are shown 
in the lower panels. Consecutive snapshots have a time difference of 20 and
are shifted in the upward direction by a distance of 0.1.
We used an inhomogeneity of the form (\ref{inhomoeq1}) with $s=10$
and $\Delta g= 0.2$, and took the group-velocity $c=1.2 c_{{ crit}}$.
\label{perdyn}}
\end{figure}
\begin{figure}
\caption[]{Mixture of chaotically oscillating DW and traveling kinks in NLD
model.
\label{mix}}
\end{figure}

\begin{figure}
\caption[]{Three examples for the dynamics for $c=c_1=0$ and $c_2=-c_3$,
all for the time interval 2500. Separate sets have a time difference 
50: (a) $c_3=0.02$, an oscillatory state; 
(b) $c_3=0.2$, the domain wall becomes chaotic and nucleates a periodic state, 
that becomes stationary at longer times;
(c) $c_3=1$, also the case when the bimodal state becomes unstable; 
it is not clear
whether a stationary periodic state sets in finally or not.
\label{cglfig1}}
\end{figure}
 
\begin{figure}
\caption[]{The oscillatory domain wall: (a) the value of $\int dx |A|/<\int dx 
|A|>$ as a function of time; (b) hidden-line plot of $|A|$ around $x=-100$ 
for the first three oscillations ($t$ from 0  to 1250); the
hidden-line plots of the local wavenumbers of $A$ (c) and $B$ (d).
\label{cglfig2}}
\end{figure}

\begin{figure}
\caption[]{Two examples of a spatiotemporal chaos, for $c=c_1=c_2=0$:
(a) $c_3=0.5$, the periodic single-mode state has become chaotic, but the 
single-mode state seems rather passive, only disturbed by ingoing perturbations
generating by the fluctuating domain wall;
(b) $c_3=1$, both the single and bimodal state have become spatio-temporally
chaotic. \label{cglfig3}}
\end{figure}

\include{figures}
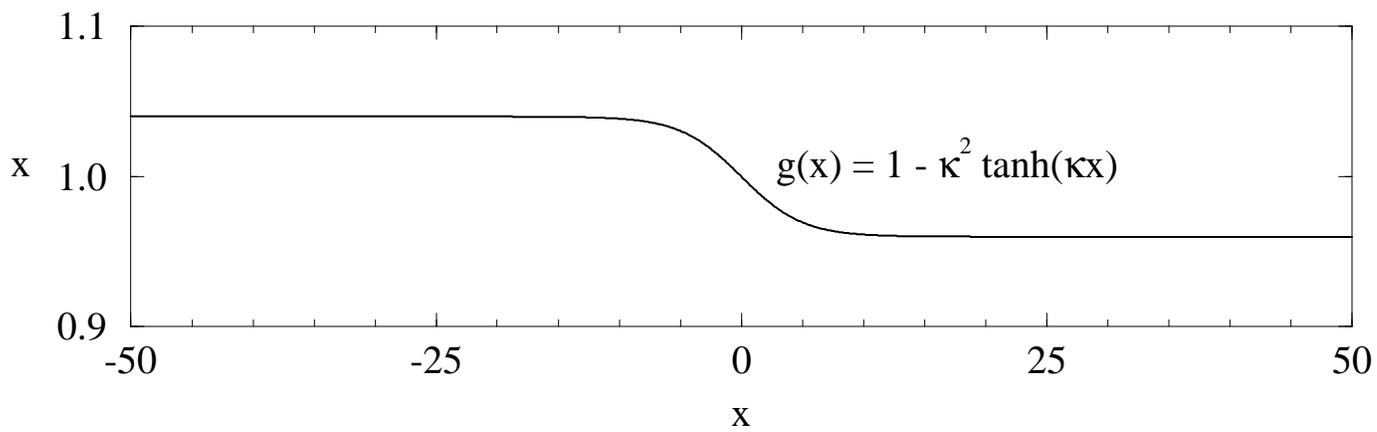
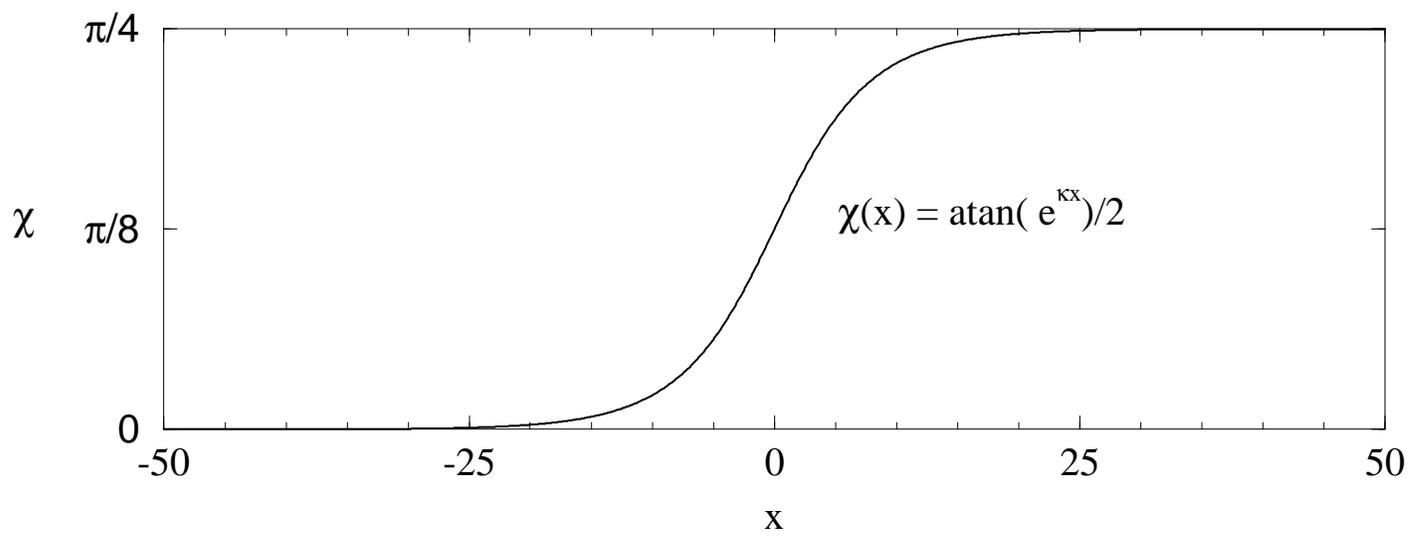
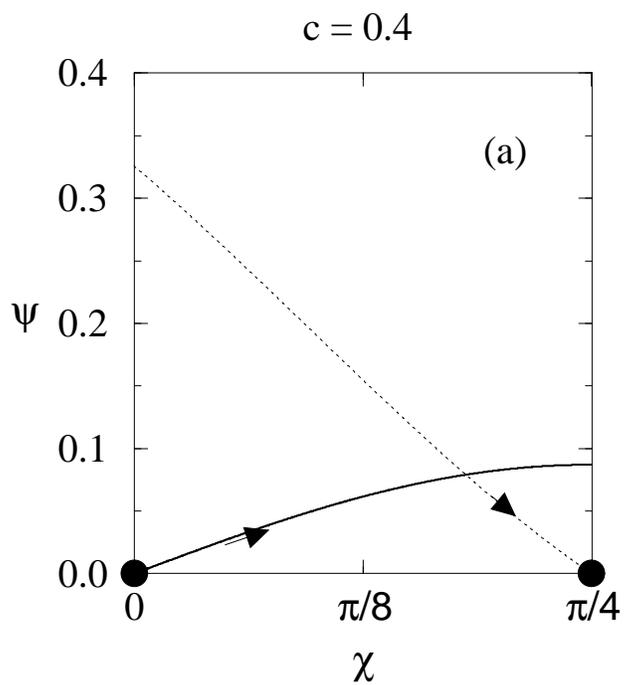
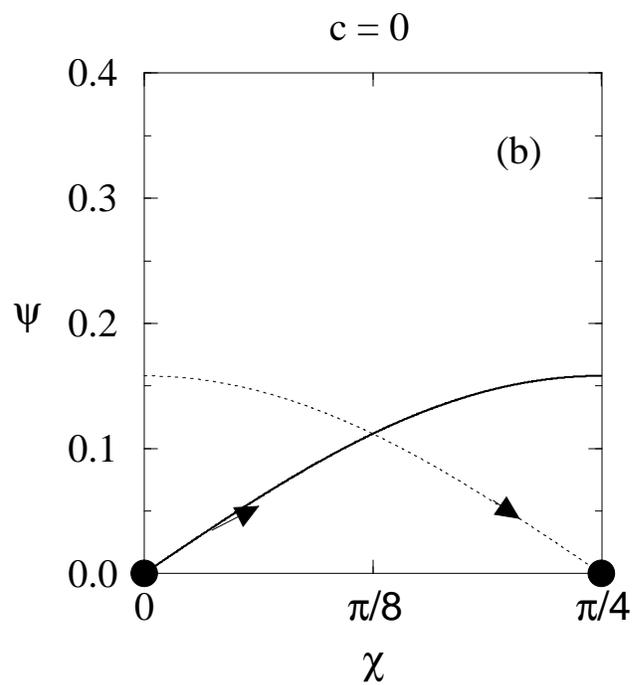
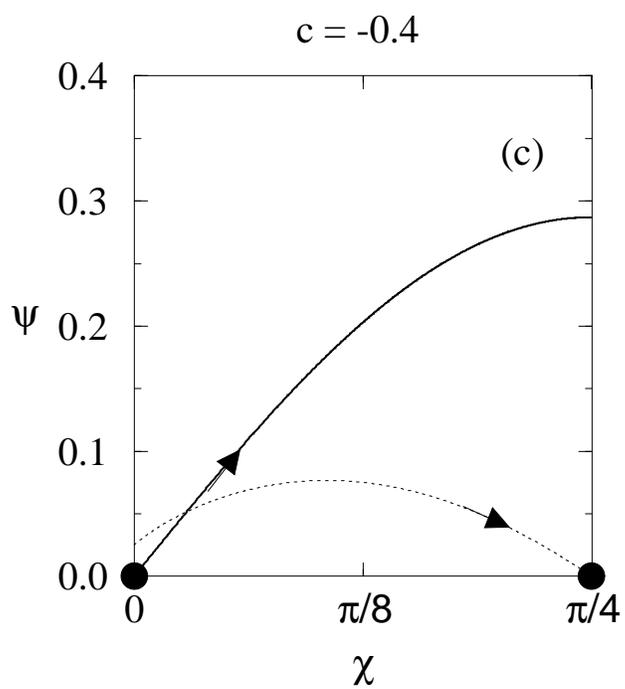
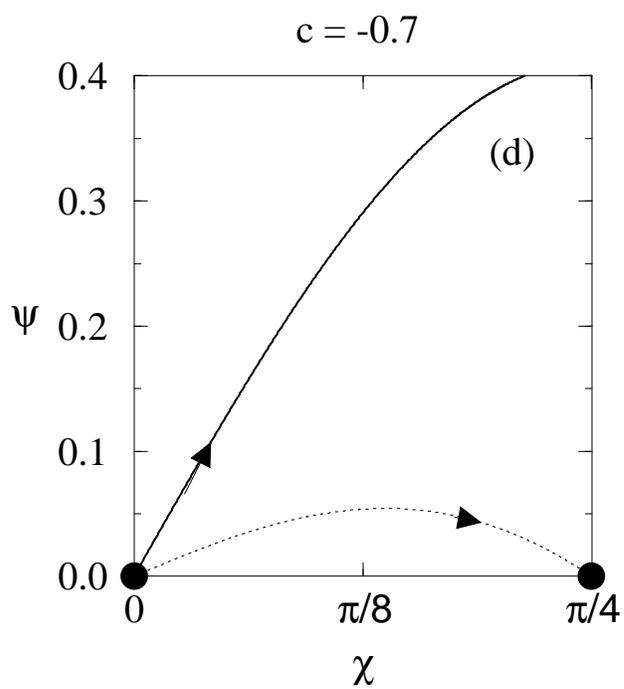
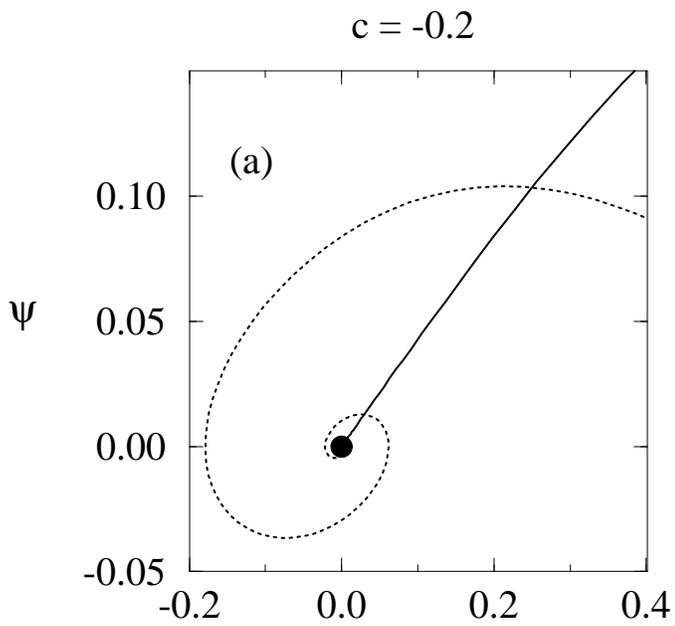
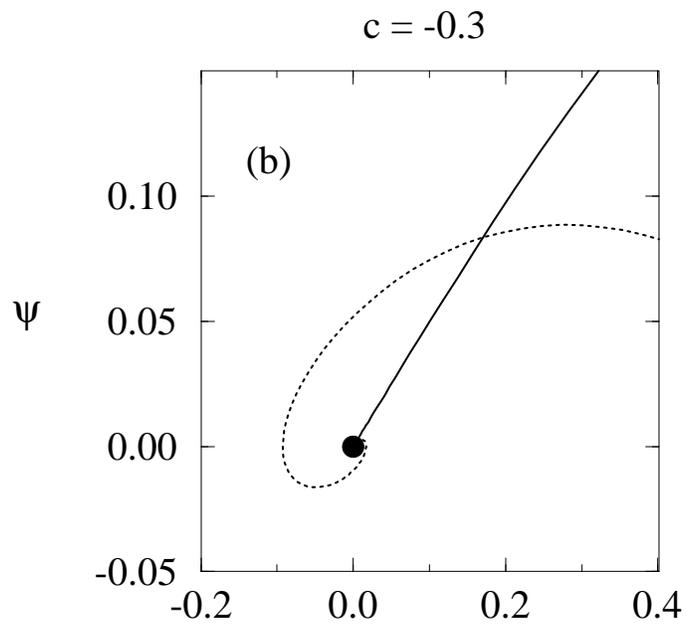
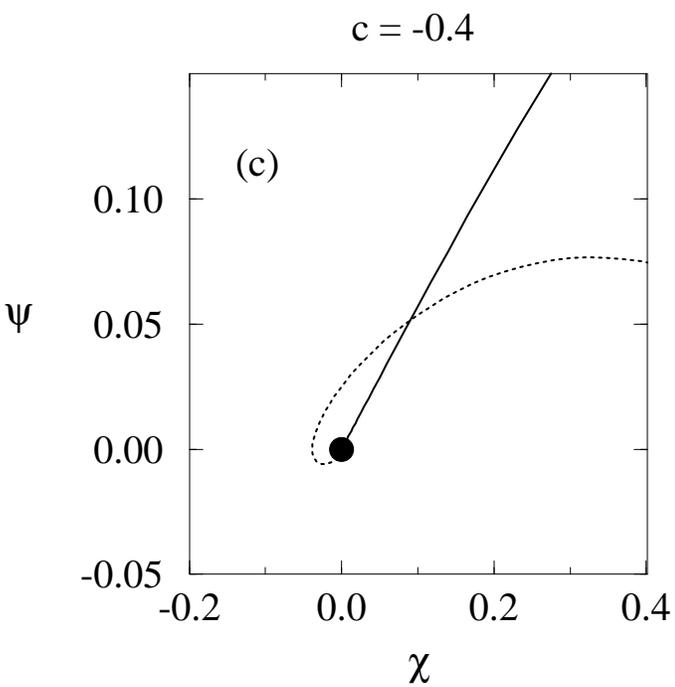
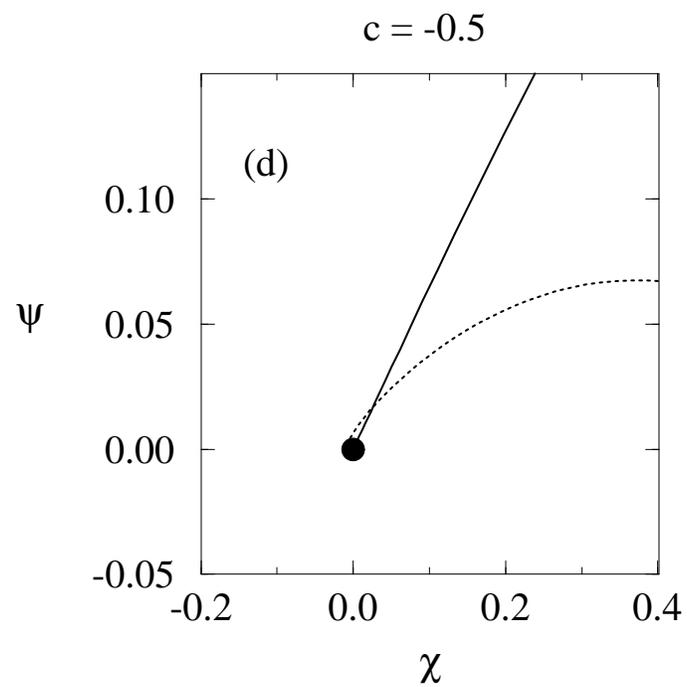
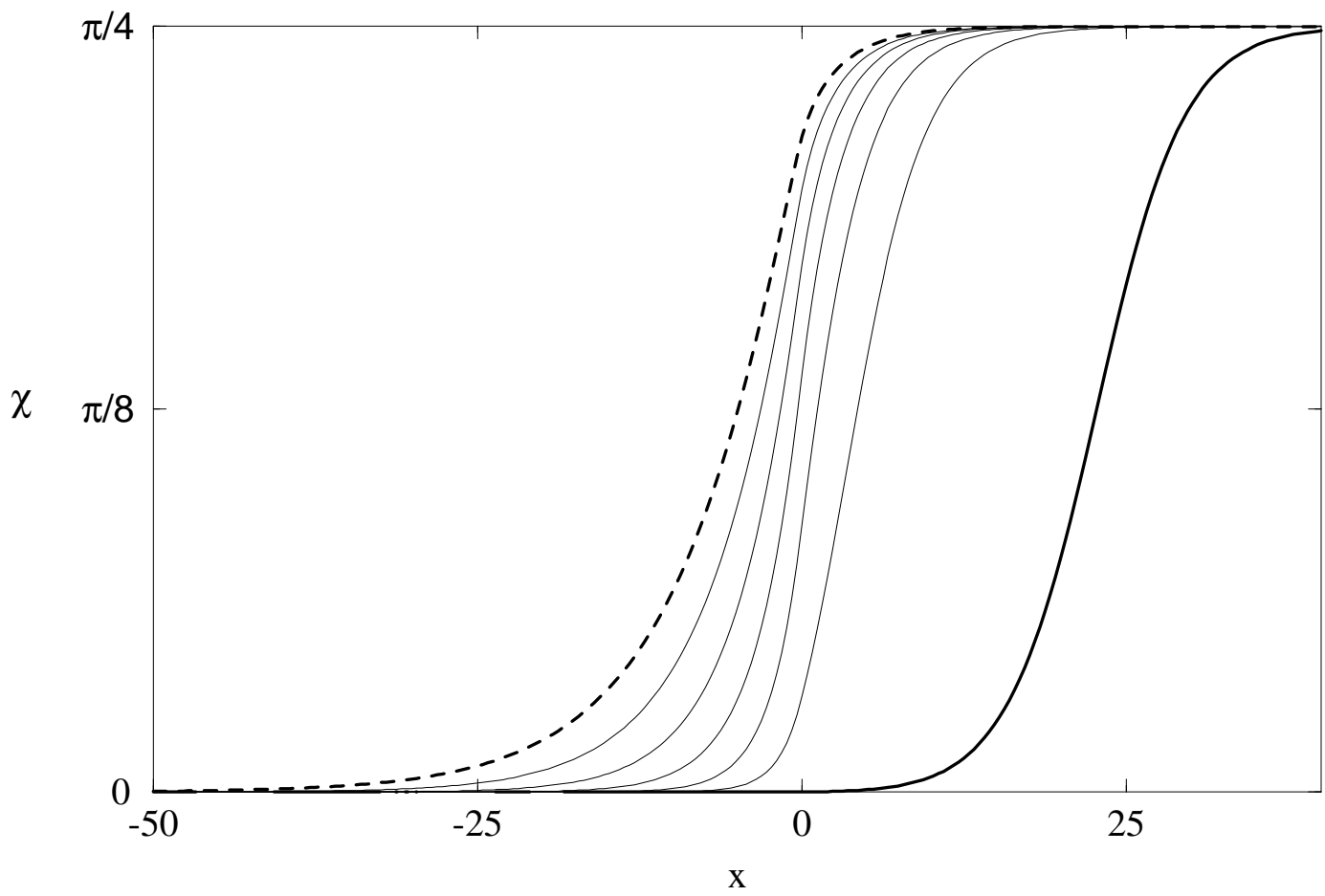
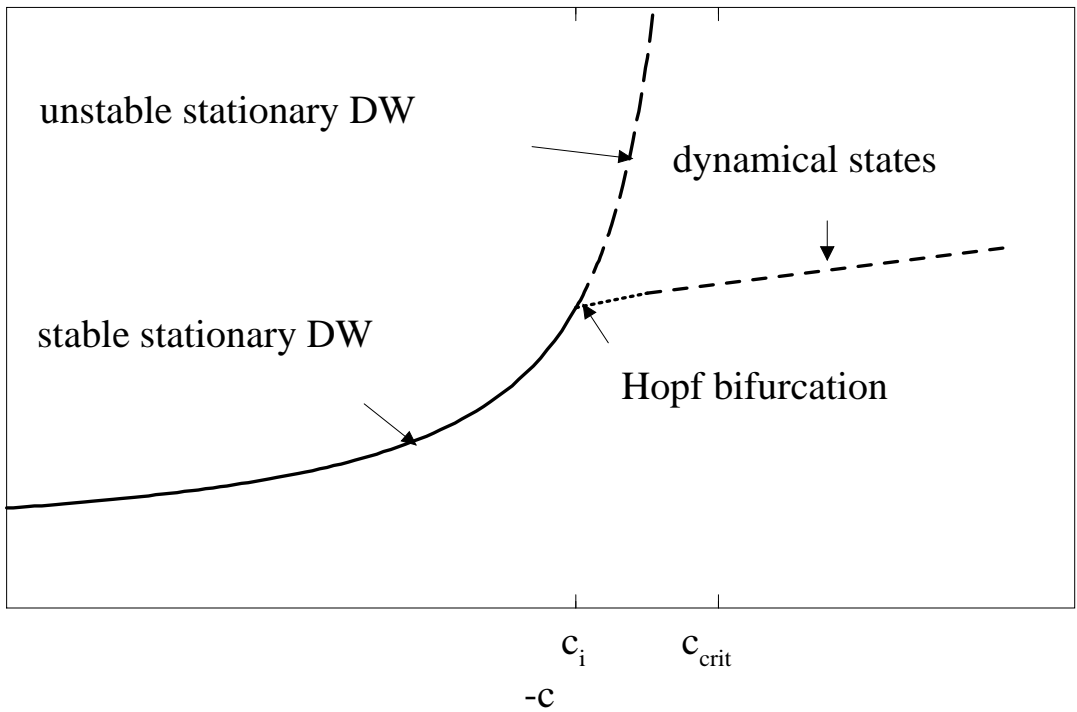
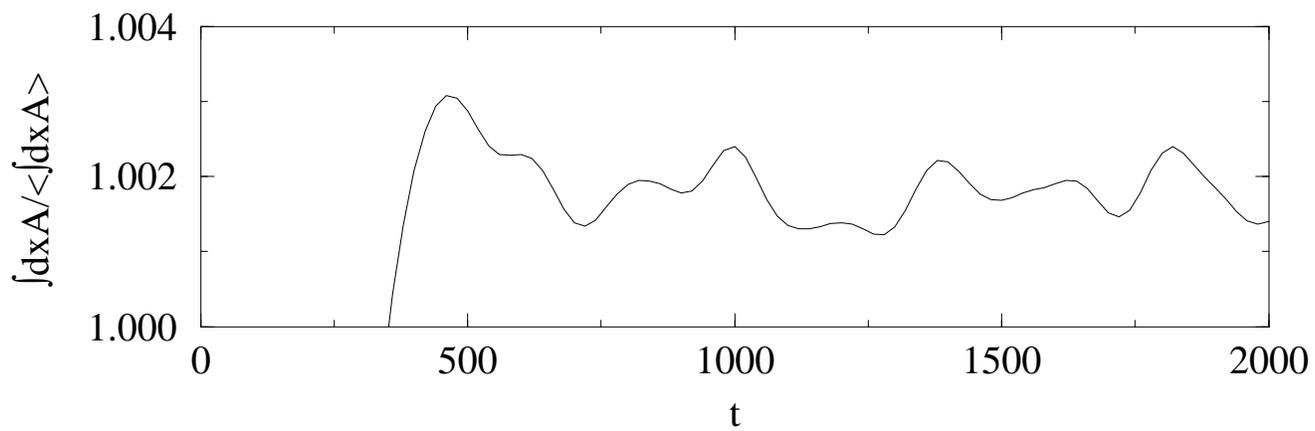
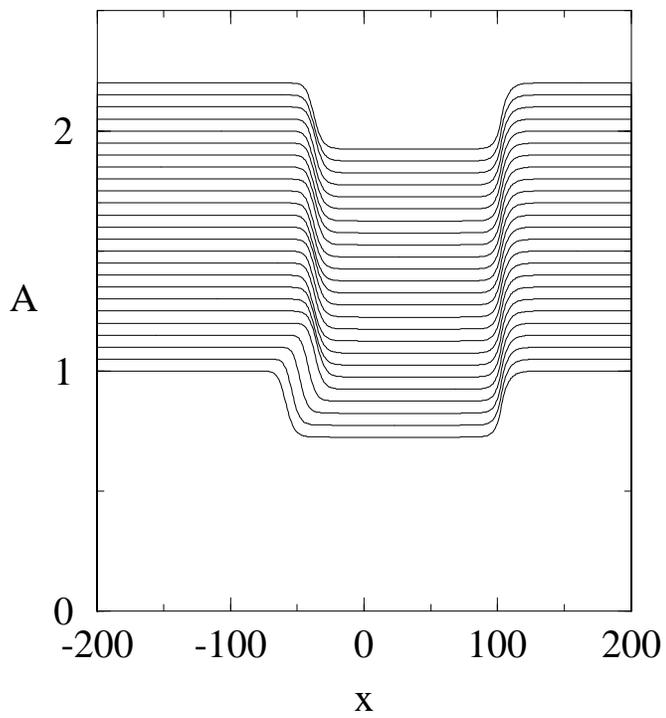
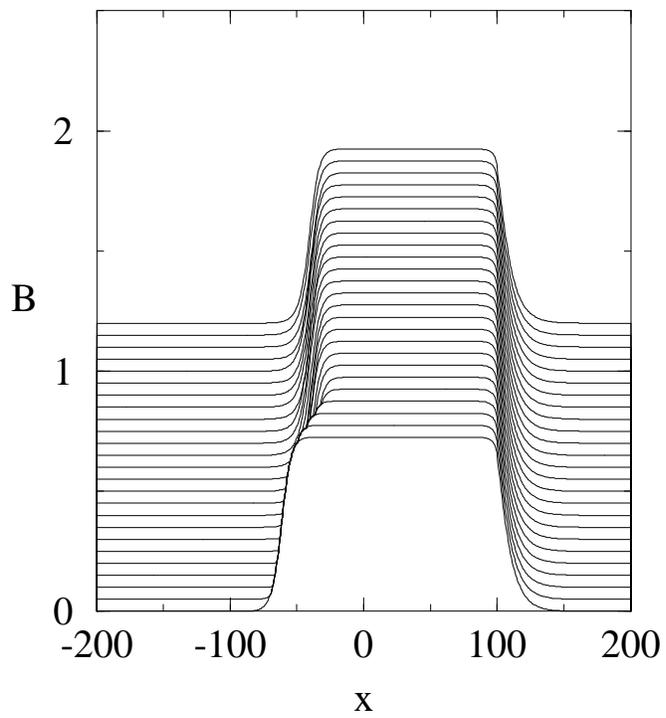
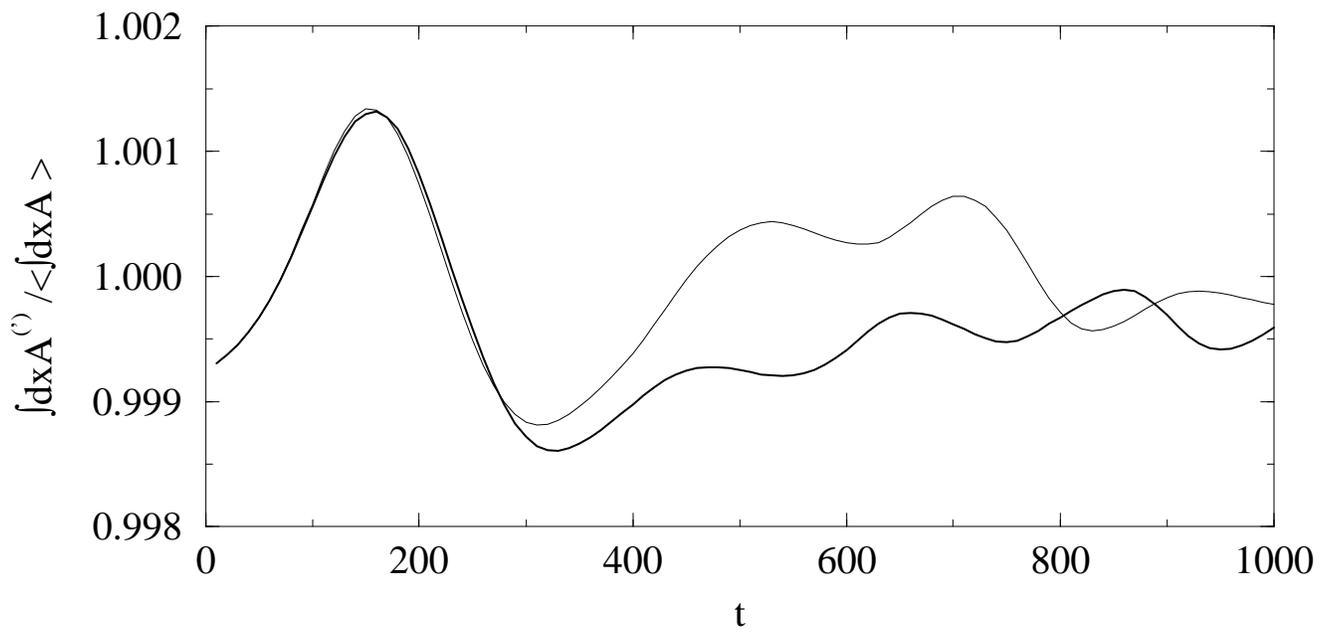
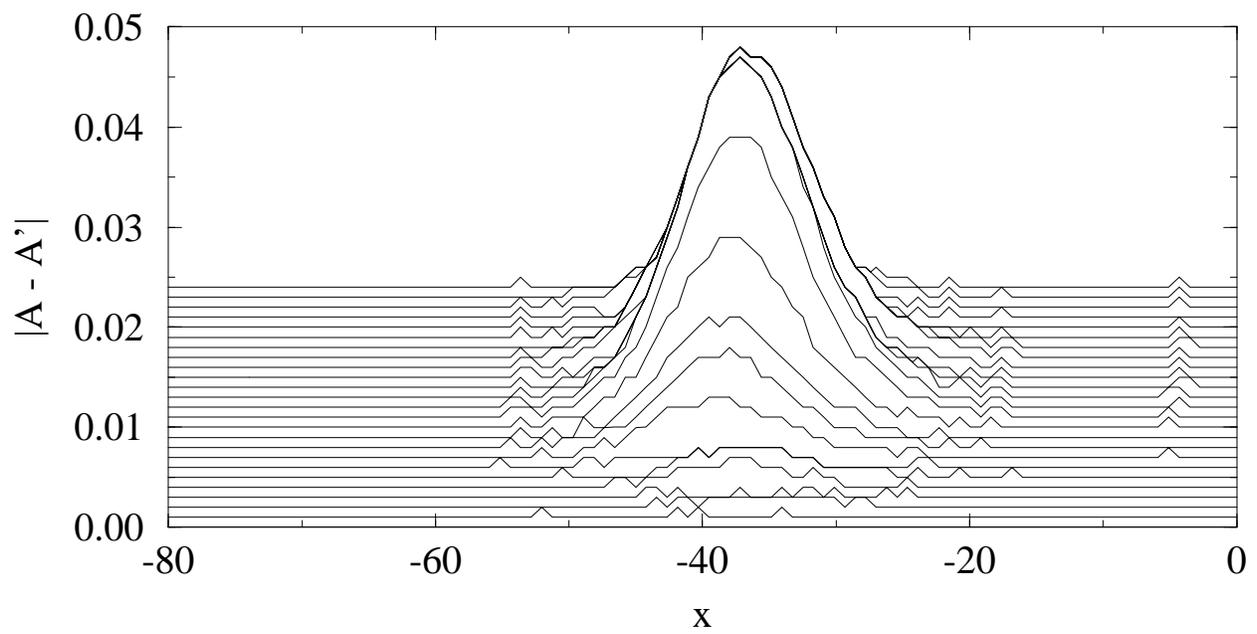
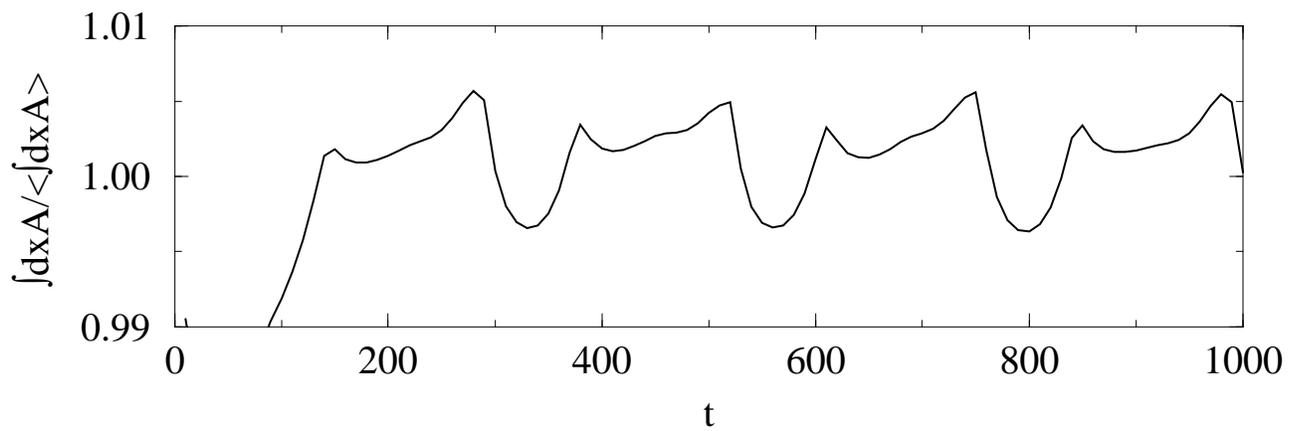
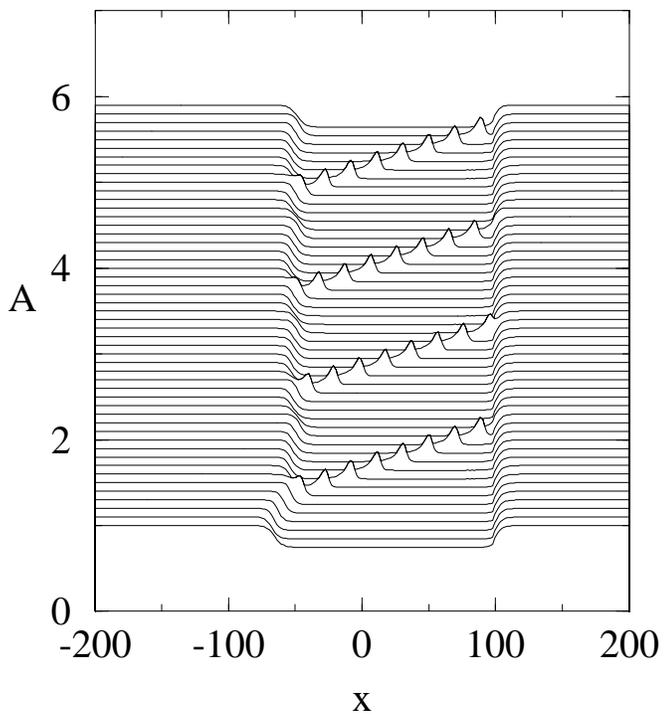
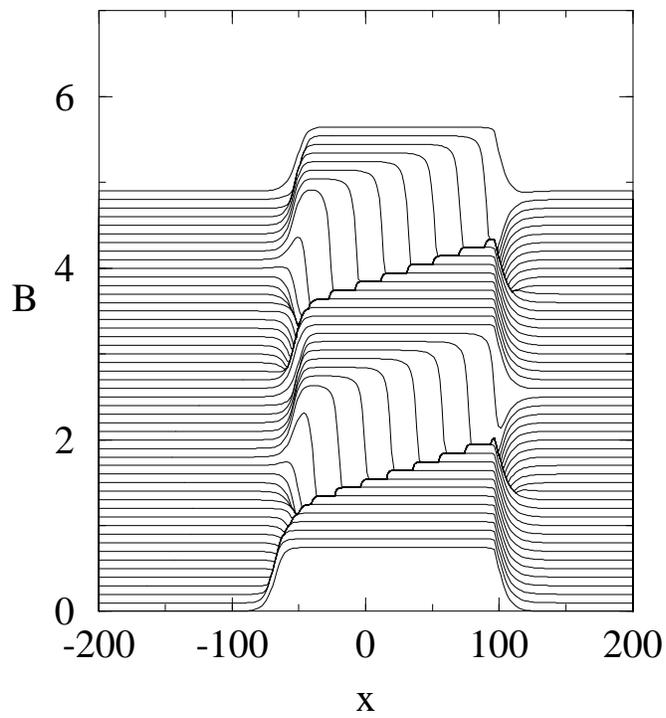
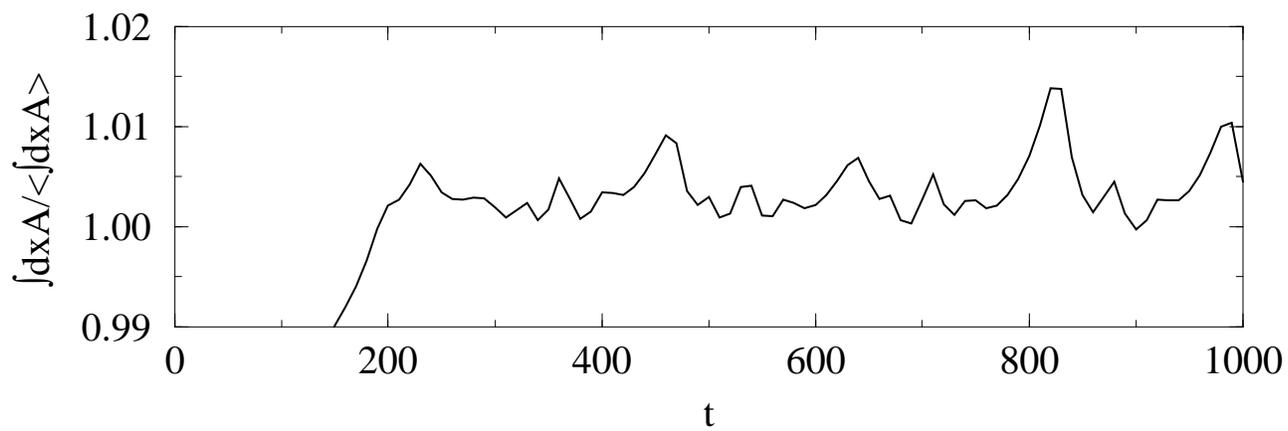
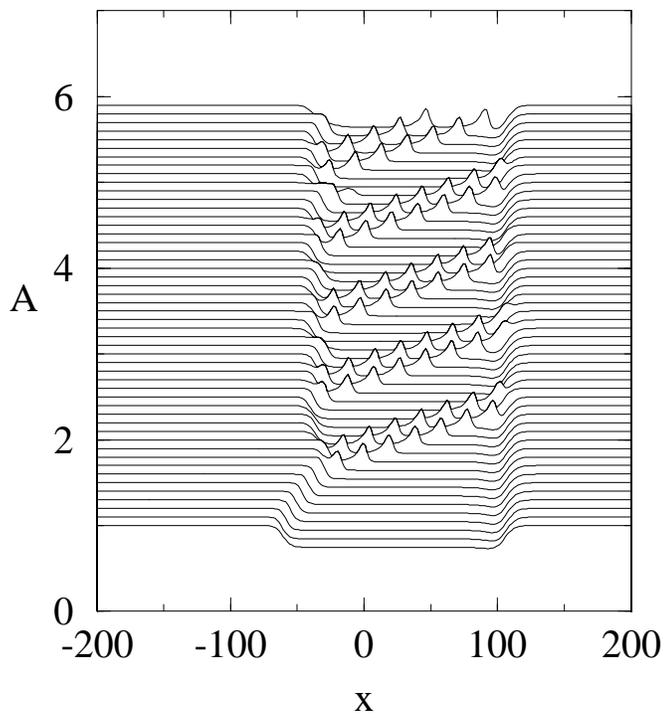
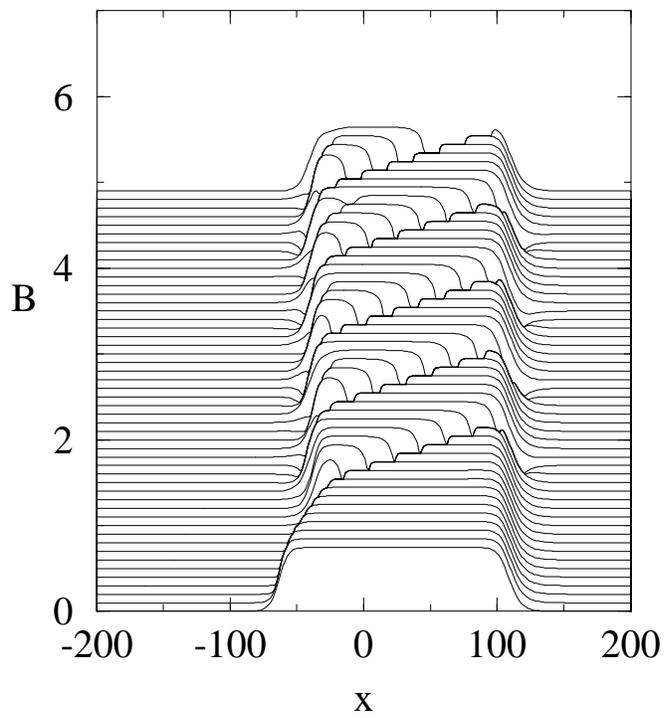
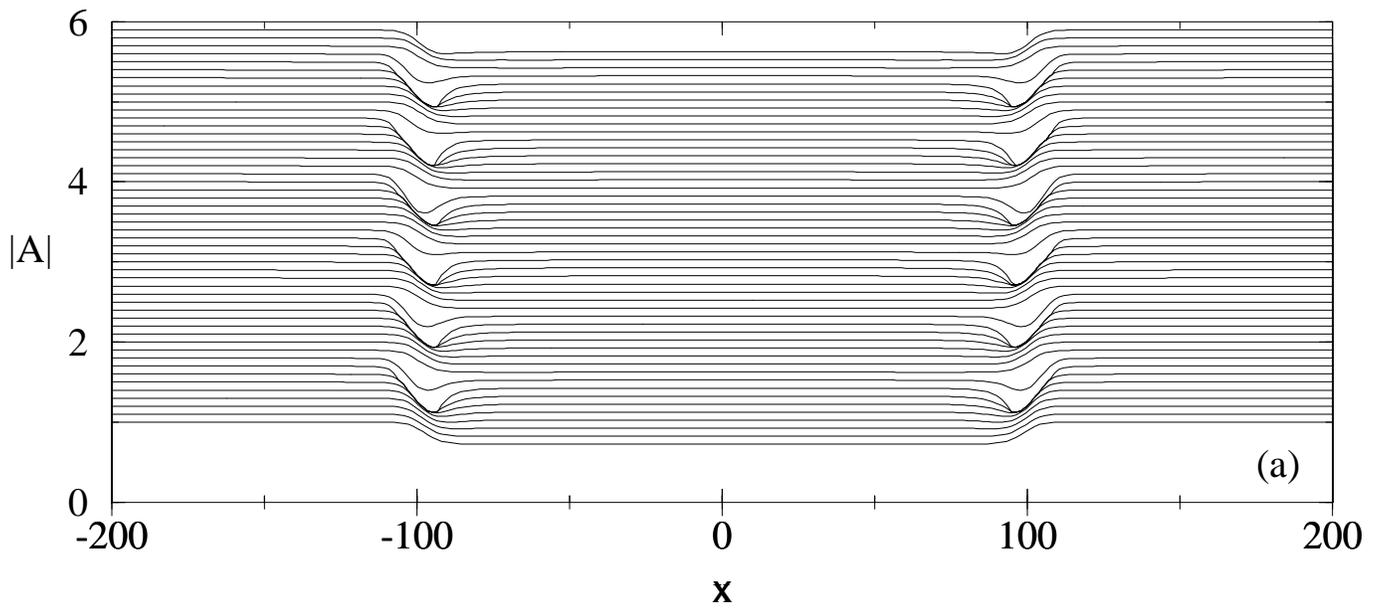
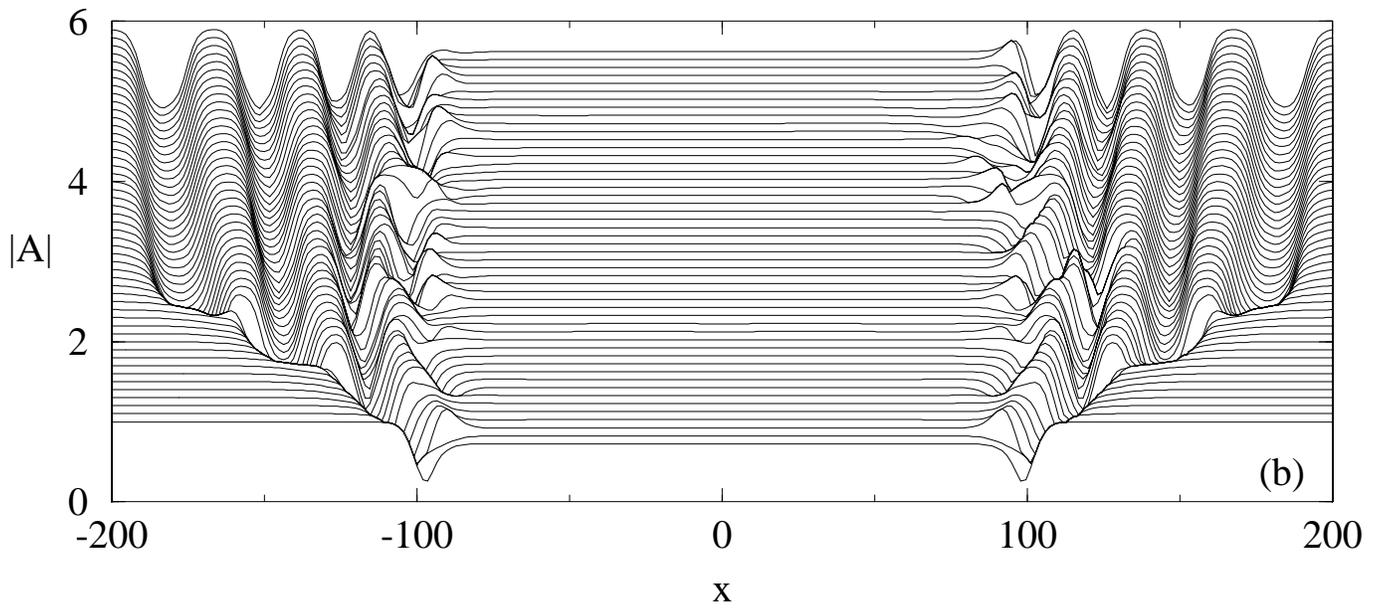
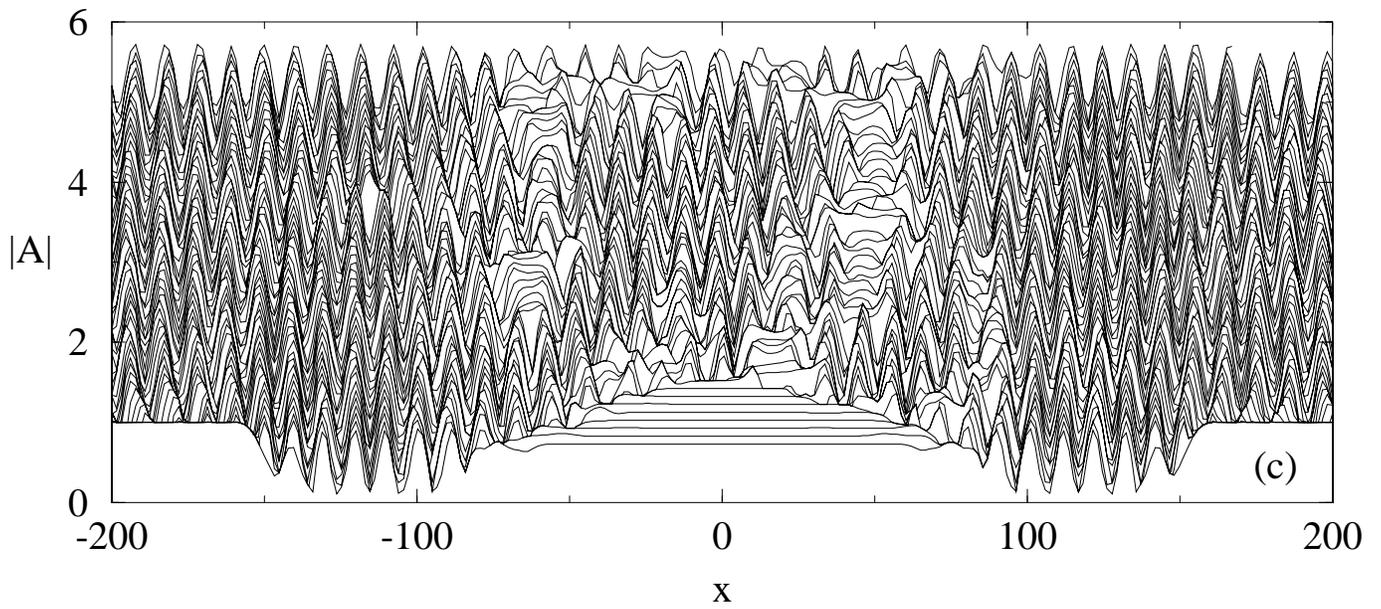
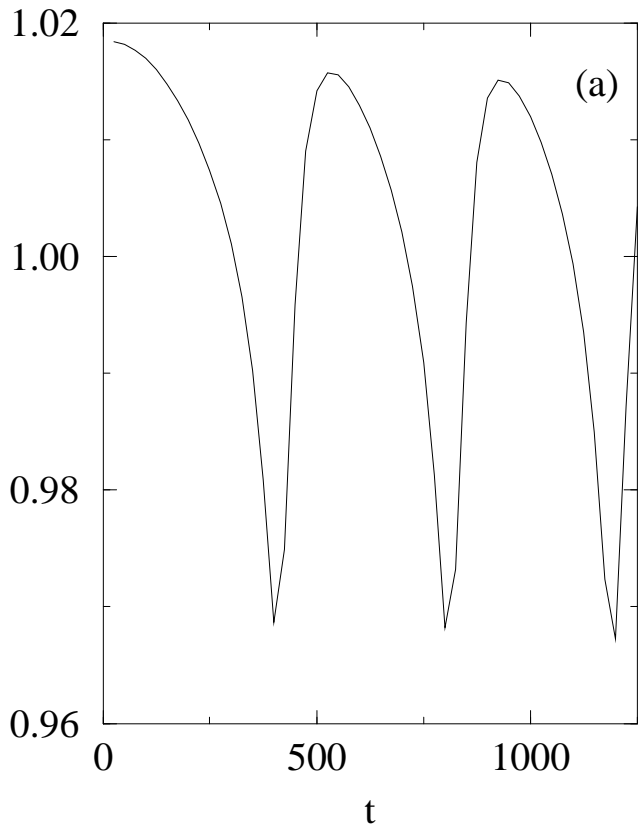
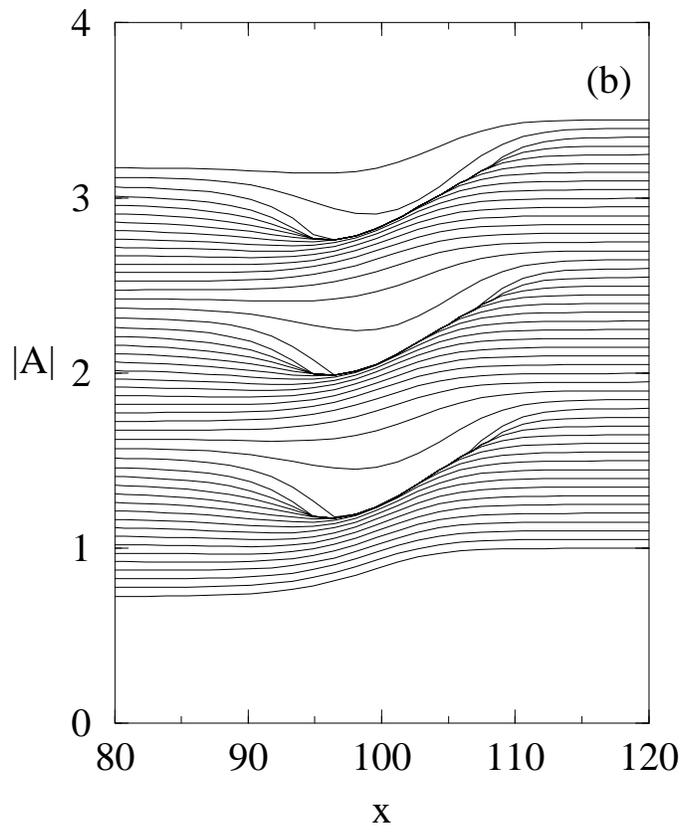
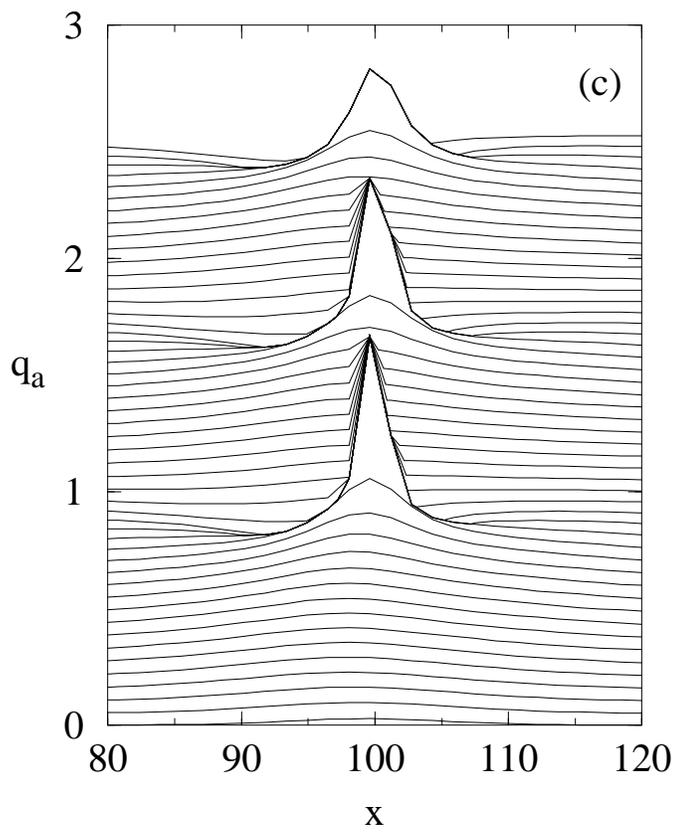
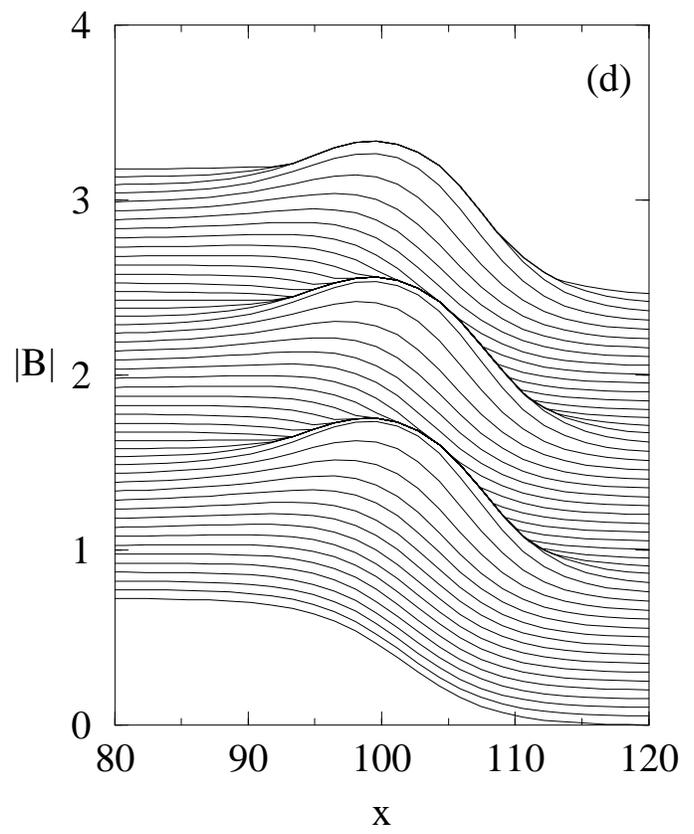
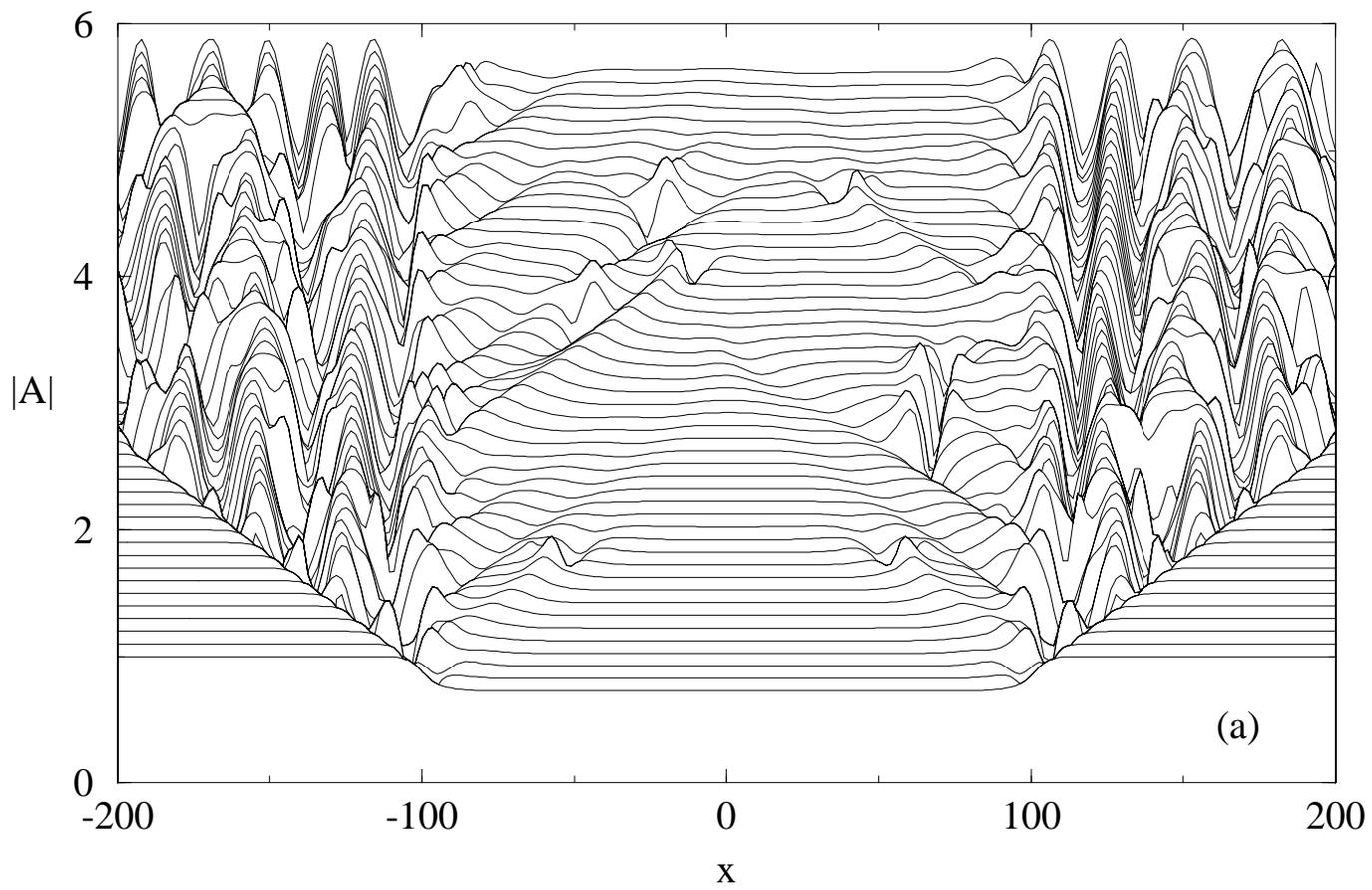
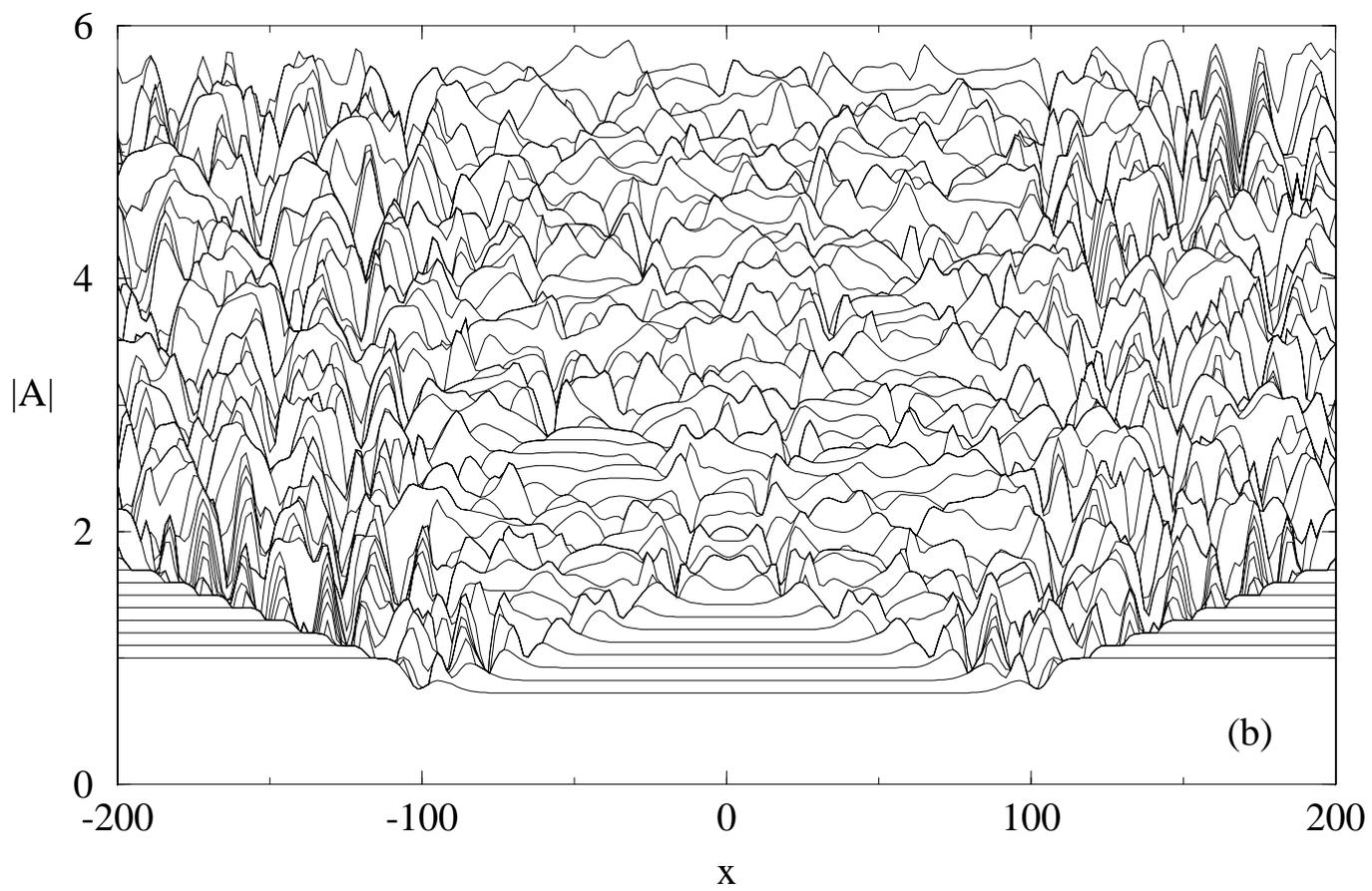
\end{document}